\documentclass[%
 reprint,
superscriptaddress,
 amsmath,amssymb,
 aps,
]{revtex4-2}

\usepackage{graphicx}
\usepackage{dcolumn}
\usepackage{bm}
\usepackage{algorithm2e}

\usepackage{todonotes}

\usepackage{hyperref}
\usepackage{amsthm}
\usepackage{comment}
\usepackage{enumitem}
\usepackage{mathtools}
\usepackage{natbib}
\usepackage{amsfonts}
\usepackage{amssymb}

\newtheorem{claim}{Claim}

\newcommand{\tr}{\text{Tr}}

\newcommand{\mg}{\mathcal G}

\newcommand{\me}{\mathcal E}
\newcommand{\mv}{\mathcal V}
\newcommand{\pp}{\mathcal P}
\newcommand{\ppq}{\mathcal Q}

\newcommand{\smin}{\setminus}
\newcommand{\nn}{\mathcal{NN}}
\newcommand{\equi}[1]{%
  #1/{\sim}%
  }
  \newcommand{\bigo}{\mathcal{O}}

\usepackage {tikz}
\usetikzlibrary {positioning}
\usetikzlibrary{decorations.pathreplacing}

\tikzset{small dot/.style={fill=black,circle,scale=1.5}}
\tikzset{small dot 1/.style={fill=red!40,circle,scale=1.5}}
\tikzset{small dot 2/.style={fill=blue!40,circle,scale=1.5}}
\tikzset{small dot 3/.style={fill=green!40,circle,scale=1.5}}
\tikzset{small dot 4/.style={fill=red!70,circle,scale=1.5}}
\tikzset{small dot 5/.style={fill=blue!70,circle,scale=1.5}}
\tikzset{small dot 6/.style={fill=green!70,circle,scale=1.5}}

\usepackage{subcaption}
\usepackage{bbm}

\usepackage{lmodern}
\usepackage{physics}

\usetikzlibrary{fit}

\begin{document}

\title{Belief propagation for networks with loops:\\ The neighborhoods-intersections-based method}

\author{Pedro Hack}
\email{pedro.hack@dlr.de}
\affiliation{German Aerospace Center, Germany}
\affiliation{Technical University of Munich, Germany}

\begin{abstract}
In order to diminish the damaging effect of loops on belief propagation (BP), the first explicit version of generalized BP for networks, the KCN-method, was recently introduced. Despite its success, the KCN-method spends computational resources inefficiently. Such inefficiencies can quickly turn the exact application of the method unfeasible, since its time complexity increases exponentially with them. This affects for instance tree networks, for which, despite not offering any accuracy advantage with respect to BP, the time complexity of the KCN-method grows exponentially with the nodes' degree.
To avoid these issues, we introduce here a new generalized BP scheme, the NIB-method, which only spends computational resources provided they are needed in order to account for correlations in the network. In fact, we show that, given a network with only short loops, the NIB-method is exact and optimal, and we characterize its time complexity reduction with respect to the KCN-method. If long loops are also present, both methods become approximate. In this scenario, we discuss the relation between the methods and
we show how to interpolate between them, obtaining a richer family of generalized BP algorithms that trade accuracy for complexity. Lastly, we find a good agreement between the (approximate) KCN and NIB methods when computing the partition function for two artificial networks.
\end{abstract}

\maketitle

\section{Introduction}

Several disciplines rely on a graph or  \textbf{network} $\mg = (\mv,\me)$, that is, a set of points or nodes $\mv$ together with a set of edges $\me$ between pairs of points, in order to study the collective behaviour of a set of individual entities (associated with the points) that interact in a pairwise fashion (associated with the edges) with each other. Networks, together with their generalization to higher order interactions among objects (i.e. \textbf{graphical models}) are used, for instance, in the study of disease spread, condensed matter physics, and both classical and quantum error correction \cite{newman2023message,bridgeman2017hand,richardson2008modern,ferris2014tensor}. 

Given the large number of interactions in several systems of interest, extracting information is a challenging task. One of the most prominent algorithms to address this problem is BP \cite{mezard2009information,pearl1988probabilistic}. While BP is computationally non-expensive, its performance decreases when facing loops. Among the proposals to cope with this, generalized BP \cite{yedidia2000generalized,welling2004choice} is based on the idea of computing correlations precisely within certain regions and then communicating this information to nearby regions in a similar fashion to how BP operates, achieving information transmission at a global scale by successively applying local communications which change, at each time step, according to the information each region received at the previous step. The lack of explicit constructions in generalized BP, together with the need for non-trivial preprocessing, led to the development of the \textbf{KCN-method} for networks \cite{kirkley2021belief,cantwell2019message}, which continues to be the state-of-the-art approach \cite{newman2023message,qian2024message}.

The KCN-method introduces a set of neighborhoods $\{N_i\}_{i \in \mv}$, on per node in the network, which are explicitly constructed in order to take care of the \textbf{short} loops around each node. Given two nodes which belong to a short loop $i,j \in \mv$, the algorithm sends a message from $i$ to $j$, and one from $j$ to $i$, which essentially communicate the non-overlapping information from $N_i$ to $N_j$ and vice versa. The computation of such non-overlapping information is performed, however, in an inefficient way in the KCN-method. By this we mean that the neighborhoods include uncorrelated pairs of nodes and, since such a lack of correlation is not taken into account when sending messages, the complexity is unnecessarily large. (This is particularly harmful in an exact implementation of the KCN-method, since the complexity grows exponentially with the number of nodes.) We can draw a direct connection between this issue and the introduction of standard BP, which was developed precisely by realizing that, given the simple correlations in a tree network, one can deal with several pairs of variables separately (hence, use less computational resources) without reducing the accuracy \cite{pearl2000models,mezard2009information}. In order to mimic the insight from BP in the context of non-tree graphs more faithfully than the KCN-method, we introduce here the \textbf{NIB-method} and study its properties, emphasizing how they compare to the KCN-method.

\subsection{Outline}

The paper is organized as follows.
We begin by fixing the setup in  Section \ref{sec: definitions}, and by introducing in Section \ref{sec: network bp} network BP, a specific network variant of standard BP that we will use here to benchmark the use of computational resources for tree-like approximations of networks. After recalling the KCN-method in Section \ref{sec: kcn-method}, we go on to highlight its inefficiency by comparing it to network BP in Section \ref{sec: kcn-method vs bp}. As a first step to avoiding such inefficiency, in Section \ref{sec: bounded nib}, we introduce the (exact) NIB-method for networks where loops are bounded in a certain sense. We characterize the instances where the NIB-method improves upon the KCN-method, and go on to show that the NIB-method is essentially optimal in this scenario. We continue in Section \ref{sec: unbounded nib}, where we eliminate the loop bound from the previous section and introduce the general (approximate) NIB-method. We compare the NIB-method to network BP in Section \ref{sec: nib vs bp}, and we show how to interpolate between the KCN and NIB methods in Section \ref{sec: interpolation}. We conclude by numerically comparing the KCN and NIB methods in Section \ref{sec: numerics}.

 \section{Networks}
 \label{sec: definitions}
 
\begin{figure}
\begin{subfigure}{0.5\textwidth}
 \centering
\begin{tikzpicture}[scale=0.5, every node/.style={transform shape}]

 \node[small dot] at (6,6) (TUU1) {};
  \node[small dot] at (4,9) (TUU2) {};

 \node[small dot] at (-2,6) (TU1) {};
\node[small dot 1] at (2,6) (TU2) {};

\node[small dot 2] at (0,3) (TM2) {};
\node[small dot] at (-4,3) (TM1) {};
\node[small dot] at (4,3) (TM3) {};

\node[small dot 4] at (-2,0) (TD1) {};
\node[small dot] at (2,0) (TD2) {};

 \node[small dot] at (-6,0) (TD0) {};
  \node[small dot] at (6,0) (TD3) {};
  
  \node[small dot] at (-2,-3) (TDD1) {};

\path[draw,thick,-]
(TD1) edge node {} (TM1)
 (TM2) edge node {} (TD1)
 (TM2) edge node {} (TM1)
 (TD1) edge node {} (TD0)
 (TM1) edge node {} (TD0)
 (TM2) edge node {} (TM3)
    (TM2) edge node {} (TD2)
    (TD2) edge node {} (TM3)
     (TM3) edge node {} (TD3)
    (TD2) edge node {} (TD3)
    (TU1) edge node {} (TM2)
    (TU2) edge node {} (TU1)
    (TU2) edge node {} (TM2)
    (TU2) edge node {} (TUU1)
    (TU2) edge node {} (TUU2)
    (TUU2) edge node {} (TUU1)
    (TD1) edge node {} (TDD1)
    ;
    
\end{tikzpicture}
  \caption{}
  \label{fig: intersection a}
\end{subfigure}%
\\
\begin{subfigure}{0.5\textwidth}
  \centering
\begin{tikzpicture}[scale=0.5, every node/.style={transform shape}]
\path[draw=blue!40, very thick] (0,-2.15) circle[radius=2.5cm];
\path[draw=red!40, very thick] (1,1.5) circle[radius=2.5cm];
\path[draw=red!70, very thick] (-2,-4.5) circle[radius=2.5cm];

\node[small dot] at (2,3) (P1) {};
\node[small dot] at (0,0) (P2) {};
\node[small dot] at (-2,-3) (P3) {};
\node[small dot] at (2,-3) (P4) {};
\node[small dot] at (-2,-6) (P5) {};

\end{tikzpicture}
  \caption{}
  \label{fig: intersection b}
\end{subfigure}%
\caption{A network instance $\mg_0$ (a), with nodes associated to variables and edges to interactions between pairs of variables. For $\mg_0$, the NIB-method can be used with three pivots, represented by colored vertices. The NIB-associated-hypernetwork $\equi{\mg_0}$ (b), with one node for each equivalence class in the intersection quotient set $\equi{\cap}$ and one hyperedge for each pivot in $\mg_0$. The hyperedges encircle the nodes in the hypernetwork that they include and are colored like the pivot in $\mg_0$ that they are associated to.}
\label{fig: intersection}
\end{figure}
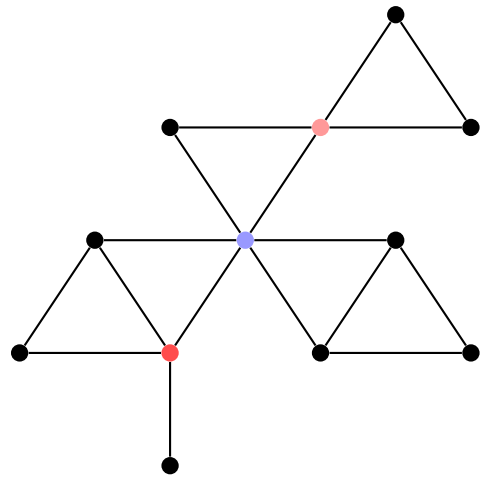
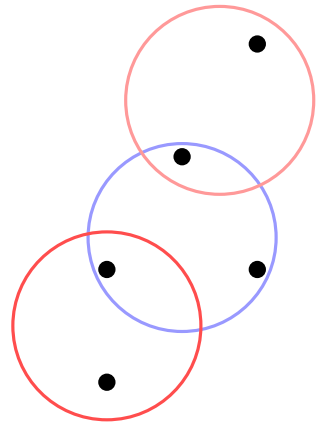

 For our purposes here, a \textbf{network} is a graph $\mg = (\mv,\me)$ where each node $i \in \mv$ is associated to a random variable $X_i$ that takes values in a finite set $X$ (which we assume to be independent of $i$ for simplicity) and, given $i,j \in \mv$, each edge $e=(i,j) \in \me$ is associated to a non-negative function $f_{i,j}: X \times X \to \mathbb R_{\geq 0}$ or \textbf{factor}. For simplicity, we assume the network to be \textbf{connected} \footnote{Otherwise, we deal with its connected components individually}. An instance of a network can be found in Figure \ref{fig: intersection a}.
 
 We think of the network as an encoding of some probability distribution that we are interested in
 \begin{equation}
 \label{eq: distribution}
     p(x_1,\dots,x_{|\mv|}) = \frac{1}{Z} \prod_{(i,j) \in \mg} f_{i,j}(x_i,x_j),
 \end{equation}
 where 
 \begin{equation*}
     Z \equiv \sum_{x_1,\dots,x_{|\mv|}} \prod_{(i,j) \in \mg} f_{i,j}(x_i,x_j)
 \end{equation*}
 is a normalization constant or \textbf{partition function} \footnote{In statistical mechanics \cite{kirkley2021belief}, it is customary to introduce networks by distinguishing between \textbf{interactions}, which correspond to the $f_{i,j}$ functions above, and \textbf{external potentials}, which are non-negative functions $f_{i}: X \to \mathbb R_{\geq 0}$ associated to each $i \in \mv$. In our simplified picture, one can assume the external potentials have been absorbed into interactions with nearest neighbors.}.
 
 In applications, networks
 may correspond to the probabilities of different spin configurations \cite{kirkley2021belief} or to the constraints satisfaction of variable configurations \cite{mezard2009information}, like the compatibility of codewords with some observed syndrome. 
 
 In order to profit from statistical mechanics' tools, it is useful to associate to each function $f_{i,j}$ some energy $E_{i,j} \equiv - \log f_{i,j}$ \footnote{In practice, the case $f_{i,j}(x,y) = 0$ for some $x,y \in X$ will not be problematic, since we will assume in our definitions that $0 \log 0 \equiv 0$.}. We can then think of \eqref{eq: distribution} as a Boltzmann distribution
 \begin{equation*}
       p(x_1,\dots,x_{|\mv|}) = \frac{1}{Z} \prod_{(i,j) \in \mg} e^{-E_{i,j}(x_i,x_j)}.
 \end{equation*}
 
 We will denote by $\mathcal{NN}_i$ the nearest neighbors of $i \in \mv$, that is, the set of nodes $j \in \mv$ that are connected to $i$ by some edge $e=(i,j)$, and use the compact trace notation $\tr(\cdot)$ ($\tr_{\smin A}(\cdot)$) to denote the sum over all variables (except those in $A$) that involve at least two functions within the parenthesis. We will refer to the number of nearest neighbors $|\mathcal{NN}_i|$ of a node $i \in \mv$ as the \textbf{degree} of $i$.
 
 \section{Standard BP on networks: Network BP}
 \label{sec: network bp}
 
 Standard BP is usually defined in the context of general graphical models and, as such, it includes two sorts of messages \cite{mezard2009information}: from factors to variables and from variables to factors. Given its restriction on factors, we can reduce this to a single type of messages between pairs of variable in the case of networks. More specifically, we define \textbf{network BP}  
 via a family of one-dimensional messages between pairs of nearest neighbors
\begin{equation*}
    \begin{split}
        &\{m^{(\text{BP}, t)}_{i \to j}\}_{i \in V, j \in \nn_i, t \geq 0}, \\
    &m^{(\text{BP}, t)}_{i \to j}: X \to \mathbb R_{\geq 0}.
    \end{split}
\end{equation*}
These messages are uniformly initialized,
and updated, for $t \geq 0$, according to the following equation:
\begin{equation*}
m_{i \to j}^{(\text{BP}, t+1)} (x_j) \propto  \tr  \biggl( f_{i,j}
\prod_{k \in \nn_i \smin \{ j \}} m_{k \to i}^{(\text{BP}, t)} \biggr),
\end{equation*}
where we omit a normalization constant that ensures the messages are probability distributions.

If we denote by 
\begin{equation*}
    \{m^{(\text{BP})}_{i \to j}\}_{i \in V, j \in \nn_i}
\end{equation*}
the set of converged messages, then we can use the equations in Appendix \ref{sec: inference network BP} to infer quantities of interest like the partition function. These equations will be exact provided the network is a tree and only approximate if that is not the case.

Let us remark that this is not how standard BP is usually simplified in the network case \cite{mezard2009information}. The reason for this is that we are here restricted to some specific type of update equations, the \textbf{single variable BP algorithms} (see Appendix \ref{sec: proof bounded claims}), which we can then scale to take into account loops in the network. Since it does not possess this scaling property, this restriction does not include the usual BP reduction to networks.
In any case, the time complexity only increases linearly with the size of the network, which is negligible compared to the exponential inefficiency in the KCN-method that we intend to address here. 
 
 \section{The KCN-method}
 \label{sec: kcn-method}
 
 The KCN-method \cite{kirkley2021belief,cantwell2019message} goes beyond network BP by computing more correlations precisely. The number of correlations that are taken into account increases with a positive integer parameter $r \geq 0$, the \textbf{loop bound}, that is fixed for each instance of the algorithm. For instance, if we take $r=0$, then the KCN-method takes into account the same correlations as standard BP.
 
 For a given loop bound $r$, the KCN-method is defined through two families of neighborhoods:
 \begin{itemize}
     \item the \textbf{primary} neighborhoods
     \begin{equation}
     \label{eq: prim neigh}
         \{N_i^{(r)}\}_{i \in \mv},
     \end{equation}
     where $N_i^{(r)}$ consists of $i$ together with its nearest neighbors $\nn_i$ and the edges joining it to them, plus both edges and nodes along paths that join two nearest neighbors of $i$; 
     \item and the family of differences between these neighborhoods 
     \begin{equation}
     \label{eq: diff neigh}
     \{N_{i \smin j}^{(r)}\}_{i \in \mv, j \in N_i^{(r)} \smin \{i\}},    
     \end{equation}
     where $N_{i \smin j}^{(r)}$ consists of node $i$ together with all the edges in $N_i^{(r)}$ which are not in $N_j^{(r)}$, and the nodes at the endpoints of such edges.
 \end{itemize}
 We will omit the superscript $r$ in the following whenever it is clear. 
 
 \begin{figure}
\begin{subfigure}{0.25\textwidth}
  \centering
 \begin{tikzpicture}[scale=0.5, every node/.style={transform shape}]

 \node[small dot 2] at (-2,6) (TU1) {};
\node[small dot 2] at (2,6) (TU2) {};

\node[small dot 2, label={[anchor=east,above=1mm, thick, font=\fontsize{18}{18}\selectfont, thick]90:\textbf{$i$}}] at (0,3) (TM2) {};
\node[small dot 2] at (-4,3) (TM1) {};
\node[small dot 2] at (4,3) (TM3) {};

\node[small dot 2] at (-2,0) (TD1) {};
\node[small dot 2] at (2,0) (TD2) {};

\path[draw,thick,blue,-]
(TD1) edge node {} (TM1)
 (TM2) edge node {} (TD1)
 (TM2) edge node {} (TM1)
 (TM2) edge node {} (TM3)
    (TM2) edge node {} (TD2)
    (TD2) edge node {} (TM3)
    (TU1) edge node {} (TM2)
    (TU2) edge node {} (TU1)
    (TU2) edge node {} (TM2)
    ;

    \end{tikzpicture}
  \caption{}
\end{subfigure}%
\begin{subfigure}{0.25\textwidth}
  \centering
 \begin{tikzpicture}[scale=0.5, every node/.style={transform shape}]

 \node[small dot 2] at (-2,6) (TU1) {};
\node[small dot 2] at (2,6) (TU2) {};

\node[small dot 2, label={[anchor=east,above=1mm, thick, font=\fontsize{18}{18}\selectfont, thick]90:\textbf{$i$}}] at (0,3) (TM2) {};
\node[small dot 2] at (-4,3) (TM1) {};
\node[small dot, , label={[anchor=east,above=1mm, thick, font=\fontsize{18}{18}\selectfont, thick]90:\textbf{$j$}}] at (4,3) (TM3) {};

\node[small dot 2] at (-2,0) (TD1) {};
\node[small dot] at (2,0) (TD2) {};

\path[draw,thick,-]
 (TM2) edge node {} (TM3)
    (TM2) edge node {} (TD2)
    (TD2) edge node {} (TM3)
    ;
    
    \path[draw,thick,blue,-]
(TD1) edge node {} (TM1)
 (TM2) edge node {} (TD1)
 (TM2) edge node {} (TM1)
    (TU1) edge node {} (TM2)
    (TU2) edge node {} (TU1)
    (TU2) edge node {} (TM2)
    ;

    \end{tikzpicture}
  \caption{}
\end{subfigure}%
\\
\begin{subfigure}{0.5\textwidth}
  \centering
 \begin{tikzpicture}[scale=0.5, every node/.style={transform shape}]

 \node[small dot] at (-2,6) (TU1) {};
\node[small dot] at (2,6) (TU2) {};

\node[small dot 2, label={[anchor=east,above=1mm, thick, font=\fontsize{18}{18}\selectfont, thick]90:\textbf{$i$}}] at (0,3) (TM2) {};
\node[small dot] at (-4,3) (TM1) {};
\node[small dot 2, label={[anchor=east,above=1mm, thick, font=\fontsize{18}{18}\selectfont, thick]90:\textbf{$j$}}] at (4,3) (TM3) {};

\node[small dot] at (-2,0) (TD1) {};
\node[small dot 2] at (2,0) (TD2) {};

\path[draw,thick,blue,-]
 (TM2) edge node {} (TM3)
    (TM2) edge node {} (TD2)
    (TD2) edge node {} (TM3)
    ;
    
    \path[draw,thick,-]
(TD1) edge node {} (TM1)
 (TM2) edge node {} (TD1)
 (TM2) edge node {} (TM1)
    (TU1) edge node {} (TM2)
    (TU2) edge node {} (TU1)
    (TU2) edge node {} (TM2)
    ;

    \end{tikzpicture}
  \caption{}
\end{subfigure}%
\caption{The sorts of neighborhoods we consider. In blue, the primary neighborhood $N_i^{(1)}$ (a), the difference neighborhood $N_{i \smin j}^{(1)}$ (b), and the intersection neighborhood $N_{i \cap j}^{(1)}$ (c).}
\label{fig: different factor graphs}
\end{figure}
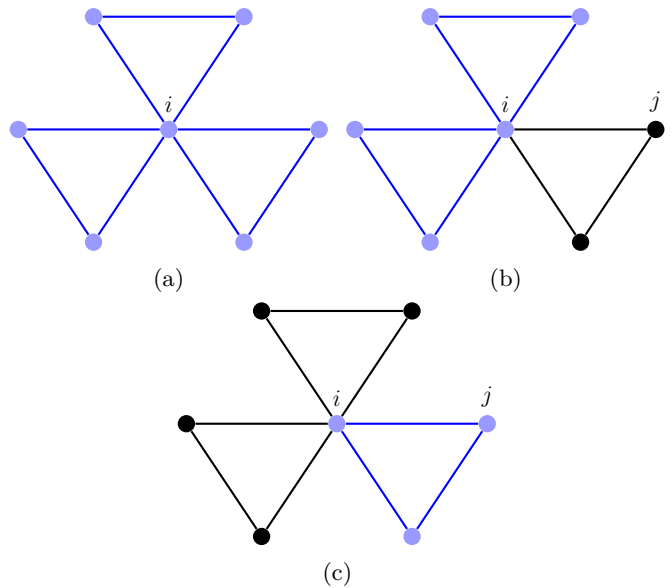
 
The \textbf{KCN-method}, or \textbf{$r$-KCN-method} to be more specific, is defined via a family of one-dimensional messages between pairs of nodes that belong to each other's primary neighborhood
\begin{equation*}
    \begin{split}
        &\{m^{(\text{KCN}, t)}_{i \to j}\}_{i \in \mv, j \in N_i \smin \{i\}, t \geq 0}, \\
    &m^{(\text{KCN}, t)}_{i \to j}: X \to \mathbb R_{\geq 0}.
    \end{split}
\end{equation*}
These messages are uniformly initialized,
and updated, for $t \geq 0$, according to the following equation:
\begin{equation}
\label{eq: update kcn}
m_{i \to j}^{(\text{KCN}, t+1)} (x_i) \propto  \tr_{\smin x_i}  \left( S_{i \smin j}
\prod_{k \in N_{i \smin j} \smin \{ i \}} m_{k \to i}^{(\text{KCN}, t)} \right),
\end{equation}
where $S_{i \smin j} \equiv \prod_{(k,q) \in N_{i \smin j}} f_{k,q}$ stands for the product of the factors within the neighborhood difference $N_{i \smin j}$ , and we omit a normalization constant that ensures the messages are probability distributions.

 If we denote by 
\begin{equation*}
    \{m^{(\text{KCN})}_{i \to j}\}_{i \in \mv, j \in N_i}
\end{equation*}
the set of converged messages, then we can use the equations in Appendix \ref{sec: kcn inference} to infer quantities of interest like the partition function. These equations will be exact provided, for each $i \in \mv$, all loops around $i$ are contained within $N_i^{(r)}$. If that is the case, we say that the loop bound $r$ is \textbf{fulfilled}, and that the network $\mg$ has \textbf{$r$-bounded} loops. (\textbf{$r$-unbounded} loops if that is not the case.) Moreover, provided $r$ is the smallest loop bound that is fulfilled, we say $r$ is \textbf{sharp}. If $r$ is not fulfilled, the the inference equations in Appendix \ref{sec: kcn inference} are only approximate.

 For a given network $\mg_0$, we say the KCN-method is
 \textbf{unfeasible} provided the time complexity of the update \eqref{eq: update kcn} or inference (Appendix \ref{sec: kcn inference}) equations is too large to be directly computed.
 
  \section{Computational resources in the KCN-method}
  \label{sec: kcn-method vs bp}
  
 Intuitively, a generalized BP method should only be unfeasible provided the variables in the network are correlated too strongly (i.e. the network has several loops), and the method is using computational power in order to take such correlations into account. However, the suboptimal use of computational resources can \textbf{artificially} turn the KCN-method unfeasible. The best way to illustrate this is by showing that the KCN-method is unfeasible for some networks with minimal variable correlations, that is, trees:
 
  \begin{claim}[$0$-KCN vs. network BP]
  \label{claim: kcn vs bp}
  The time complexity of the $0$-KCN-method grows exponentially with the maximal node degree of the network without providing any accuracy advantage with respect to network BP.
  \end{claim}
  
 Note that Claim \ref{claim: kcn vs bp}, whose proof we delay until Appendix \ref{sec: other inference}, does not only address tree networks but any network that is treated as a tree by the KCN-method. Intuitively, the claim makes sense since the $0$-KCN-method does not take into account more correlations than network BP. As a result of the claim, the KCN-method is unfeasible on a tree network provided there is some node whose \textbf{degree} is too large. As in the case of network BP, the nodes' degree alone (that is, without any reference to loops or correlations) will not constitute a constraint for the exact applicability of the NIB-method, which we introduce in the next section.
 
 \section{The NIB-method with $r$-bounded loops}
 \label{sec: bounded nib}

We introduced the KCN-method in Section \ref{sec: kcn-method} without distinguishing between the $r$-bounded and $r$-unbounded loops cases. In order to highlight the properties of the NIB-method, we differentiate between these cases when introducing it, addressing first the bounded case in this section, and dealing with the unbounded case in Section \ref{sec: unbounded nib} (where we also remark that they could have been introduced simultaneously as in the KCN case).
Throughout this section, we assume some loop bound $r_0$ is fulfilled.

 \subsection{The neighborhoods intersections}
 
 The NIB-method is defined through a subfamily of the intersections of the KCN primary neighborhoods \eqref{eq: prim neigh}:
 \begin{itemize}
     \item the \textbf{neighborhoods intersections}
 \begin{equation}
 \label{eq: neigh inter}
 \begin{split}
     &\{N_{i \cap j} \}_{i \in \mv, j \in N_i \smin \{i \}}, \text{ where} \\
     &N_{i \cap j} \equiv N_i \cap N_j
     \end{split}
 \end{equation}
 and $\cap$ stands for the usual set intersection operator \footnote{Note that the neighborhoods intersections do not include the intersection of \textbf{any} pair of primary neighborhoods, but only those between pairs of nodes that belong to each other's primary neighborhood.}.
 \end{itemize}
 Intuitively, $N_{i \cap j}$ is the minimal set that includes all correlations between $i$ and $j$. We will show its minimality later on. For now, it suffices to notice that it is constructed by taking into account the loops that include both $i$ and $j$, plus paths that, although maybe not part of such a loop, cannot be disregarded without loosing exactness.
 (See Claim \ref{claim: charact intersection} in Appendix \ref{sec: definition equi classes} for a characterization of $N_{i \cap j}$.)

The neighborhoods intersections fulfill an important property, which we refer to as the \textbf{equivalence class condition}:

\begin{claim}[The equivalence class condition]
 \label{claim: intersection property}
If the loop bound $r_0 \geq 0$ is fulfilled, $i \in \mv$, and $j \in \mv \cap \left( N_i^{(r_0)} \smin \{i \} \right)$, then
\begin{equation}
\label{eq: equiv class cond}
    N_{i \cap j}^{(r_0)} = N_{k \cap q}^{(r_0)}
\end{equation}
 for all $k,q \in \mv \cap N_{i \cap j}^{(r_0)}$, $k \neq q$.
\end{claim}
 We prove Claim \ref{claim: intersection property} in Appendix \ref{sec: definition equi classes}, where we also give a counterexample to show that the condition on the loop bound $r_0$ cannot be avoided \footnote{Claim \ref{claim: intersection property} was already stated in the literature \cite{kirkley2021belief}. Nonetheless, we could not find a proof for it and, since it is central for the derivation of the NIB-method, we show it in Appendix \ref{sec: definition equi classes}.}.
 
 Claim \ref{claim: intersection property} is useful since it allows us to split the family of neighborhoods intersections (hence the network) into equivalence classes such that two different classes either have no overlap or overlap on a single variable. The equivalence relation $\sim$ that we consider on the neighborhoods intersections is the following:
 \begin{equation*}
 N_{i \cap j} \sim N_{k \cap q} \text{ if and only if } N_{i \cap j} = N_{k \cap q}.    
 \end{equation*}
 This equivalence relation is associated to the \textbf{intersection quotient set}
  \begin{equation*}
     \equi{\cap} \equiv \left\{ \overline{i \cap j} \right\}_{i \in \mv, j \in N_i \smin \{i\}},
 \end{equation*}
 where $\overline{i \cap j}$ denotes the equivalence class of $N_{i \cap j}$. By reduction to the absurd \footnote{If $|\mv \cap (\overline{i \cap j} \cap \overline{k \cap q})| \geq 2$, then we take $r,s \in \mv \cap (\overline{i \cap j} \cap \overline{k \cap q})$ with $r  \neq s$ and we get, by Claim \ref{claim: intersection property}, $N_{i \cap j} = N_{r \cap s} = N_{k \cap q}$. This implies $\overline{i \cap j} = \overline{k \cap q}$, a contradiction. Moreover, if there exist $v_1,v_2 \in \mv$, $v_1 \neq v_2$, such that $e=(v_1,v_2) \in \me \cap (\overline{i \cap j} \cap \overline{k \cap q})$, then, by definition of primary and intersections neighborhoods, $v_1,v_2 \in \mv \cap (\overline{i \cap j} \cap \overline{k \cap q})$, contradicting the first statement.}, we can show that, for $\overline{i \cap j}, \overline{k \cap q} \in \equi{\cap}$ with $\overline{i \cap j} \neq \overline{k \cap q}$,
\begin{equation}
\label{eq: intersection equi classes}
\begin{split}
    &|\mv \cap \left(\overline{i \cap j} \cap \overline{k \cap q} \right)| \in \{ 0,1\} \text{, and} \\
    &\me \cap \left(\overline{i \cap j} \cap \overline{k \cap q} \right) = \emptyset.
    \end{split}
\end{equation}
We refer to the vertices that belong to at least two different equivalence classes in $\equi{\cap}$, that is, those $s \in \mv$ for which there exist $\overline{i \cap j} \neq \overline{k \cap q}$ such that $s \in \overline{i \cap j} \cap \overline{k \cap q}$, as \textbf{pivots}, and denote the set of pivots in $\mg$ by $\text{Piv}(\mg)$.

Pivots are useful in order to analyze an $r$-bounded network $\mg$, since we can associate a hypernetwork to $\mg$ through them:
\begin{itemize}
    \item The \textbf{NIB-associated-hypernetwork}
\begin{equation*}
     \equi{\mg} \equiv \left(\equi{\cap}, \{ e_s\}_{s \in \text{Piv}(\mg)} \right)
\end{equation*}
consists of the set of nodes $\equi{\cap}$ together with one hyperedge $e_s$ for each pivot $s \in \text{Piv}(\mg)$
\begin{equation*}
    e_s \equiv \{ \overline{i \cap j} \in \equi{\cap} | s \in \overline{i \cap j} \}.
\end{equation*}
\end{itemize}
The fulfillment of the loop bound implies that $\equi{\mg}$ is a \textbf{hypertree}, which is the key property that guarantees the exactness of the bounded NIB-method. We show in Figure \ref{fig: intersection a} a network with three pivots, and in Figure \ref{fig: intersection b} its NIB-associated-hypernetwork. We will denote by $\text{diam}(\equi{\mg})$ the \textbf{diameter} \cite{mezard2009information} of the NIB-associated-hypernetwork. 
 
 \subsection{Message-passing scheme}
 
The \textbf{bounded neighborhoods-intersections-based} method (\textbf{bounded NIB-method}), or \textbf{bounded $r$-NIB-method} to be specific, is defined via a family of one-dimensional messages from the equivalence classes in the intersection quotient set into their constituent nodes:
\begin{equation*}
    \begin{split}
        &\{m^{(t)}_{\overline{i \cap j} \to i}\}_{i \in V, j \in (N_i \smin \{i\})/\sim, t \geq 0}, \\
    &m^{(t)}_{\overline{i \cap j} \to i}: X \to \mathbb R_{\geq 0},
    \end{split}
\end{equation*}
where, given $j,k \in N_i \smin \{i\}$,
\begin{equation*}
    j \sim k \text{ if and only if } \overline{i \cap j} = \overline{i \cap k}.
\end{equation*}
These messages are uniformly initialized,
and updated, for $t \geq 0$, according to the following equation:
\begin{equation}
\label{eq: intersection}
m_{\overline{i \cap j} \to i}^{(t+1)} (x_i) \propto \tr_{\setminus x_i} \left( S_{i \cap j}  \prod_{k \in N_{i \cap j} \smin \{ i \}} \prod_{\overline{k \cap q} \neq \overline{i \cap j} } m_{\overline{k \cap q} \to k}^{(t)} \right),
\end{equation}
where $S_{i \cap j} \equiv \prod_{(k,q) \in N_{i \cap j}} f_{k,q}$ is the product of the factors in $N_{i \cap j}$ and we omit a normalization constant which ensures that the messages are probability distributions \footnote{If $N_i \smin N_{i \cap j} = \emptyset$, then it will become clear that we do not need to update $m_{\overline{i \cap j} \to i}^{(t)}$ and, hence, this message will remain a uniform distribution for all $t \geq 0$.}. 

 If we denote by 
\begin{equation*}
    \{m_{\overline{i \cap j} \to i}\}_{i \in V, j \in (N_i \smin \{i\})/\sim}
\end{equation*}
the set of converged messages, then we can use the equations in Appendix \ref{sec: proof inference eq} to exactly infer quantities of interest like the partition function. These equations are exact provided the loop bound is fulfilled.

Intuitively, the NIB-method exploits the fact that, provided the loop bound is fulfilled, the NIB-associated-hypernetwork $\equi{\mg}$ is a tree-like partition of $\mg$, and uses the pivots in $\mg$ to (exactly)
transport the information in each equivalence class to the others
via the BP updates in \eqref{eq: intersection}. We show in Figure \ref{fig: intersection a} a network on which the NIB-method uses three pivots.

\subsection{Comparison to other message-passing schemes}

Since the NIB-method is exact provided the loop bound is fulfilled, we can compare it solely in terms of complexity to other exact message-passing schemes in this context, like the KCN-method. We devote the rest of this section to such a comparison.

We begin by comparing the NIB-method to the KCN-method. Provided the loop bound $r_0 \geq 0$ is sharp, then one can show that, for most networks $\mg$, the $r_0$-NIB-method retains the exactness of the $r_0$-KCN-method while reducing its complexity:
\begin{claim}[Bounded NIB vs. bounded KCN]
\label{claim: NIB vs kcn}
If $r_0$ is a sharp loop bound, then the $r_0$-NIB-method has a smaller time complexity than the $r_0$-KCN-method if and only if there is no pair $i,j \in \mv$ such that $N_{i \cap j}^{(r_0)} = \mg$.
\end{claim}

We prove Claim \ref{claim: NIB vs kcn} in Appendix \ref{sec: proof bounded claims}. The statement relies on a speed-up in the inference complexity. A similar statement can be made regarding the update complexity, as we show in Appendix \ref{sec: proof bounded claims}. (See Claim \ref{claim: update speed-up nib vs kcn}.)

As a rough estimate of the complexity advantage of the NIB-method over the KCN-method, consider two simplified asymptotic scenarios:
\begin{itemize}
    \item If the complexity of inference in the KCN-method is much larger than that of updating (i.e. that of each update step times the diameter $\text{diam}(\equi{\mg})$), then the NIB-method requires $\bigo \left( |X|^{\alpha} \right)$ less time, where
    \begin{equation*}
         \alpha \equiv \text{max}_{i \in \mv} \Bigl\{ |\mv_{N_i}| \Bigr\}-\text{max}_{\overline{i \cap j} \in  \equi{\cap}} \Bigl\{ |\mv_{N_{i\cap j}}| \Bigr\}
    \end{equation*}
    and $\mv_A$ stands for the variables in $A \subseteq \mg$. The improvement is exponential for almost any graph (see Claim \ref{claim: NIB vs kcn}).
    
     \item If the complexity of updating in the KCN-method is much larger than that of inference, then the NIB-method requires $\bigo \left( |X|^{\gamma} \right)$ less time, where
    \begin{equation*}
         \gamma \equiv \text{max}_{i \in \mv, j \in N_i \smin \{i\}} \Bigl\{|\mv_{N_{i \smin j}}|\Bigr\}-\text{max}_{\overline{i \cap j} \in  \equi{\cap}} \Bigl\{|\mv_{N_{i\cap j}}|\Bigr\}.
    \end{equation*}
 The improvement is exponential for several graphs (see Claim \ref{claim: update speed-up nib vs kcn}).
\end{itemize}

Given the advantage of using the NIB-method over the KCN-method, one may ask whether further improvements are possible. This turns out to be impossible.
More specifically, if we restrict ourselves to the class of \textbf{exact single variable BP algorithms} (see Appendix \ref{sec: proof bounded claims} for a definition), then the NIB-method cannot be improved upon:

\begin{claim}[Bounded NIB optimality]
\label{claim: NIB optimal}
If $r_0$ is a sharp loop bound, then the
    $r_0$-NIB-method is \textbf{optimal}, in the sense that, among all the exact single variable BP algorithms, the $r_0$-NIB-method
    has minimal time complexity.
\end{claim}

We prove Claim \ref{claim: NIB optimal} in Appendix \ref{sec: proof bounded claims}. As a particular case, this statement assures that network BP is optimal provided the network is loopless \footnote{As discussed in Section \ref{sec: network bp}, this is the case given our limited definition of the single variable BP algorithms. If we extend the definition, linear complexity reductions are possible.}.

The reduction in complexity that the NIB-method provides can have dramatic consequences regarding the feasibility of the method:

\begin{itemize}
 \item If $r_0$ is a sharp loop bound, then there are networks where the $r_0$-KCN-method is \textbf{unfeasible} while the $r_0$-NIB-method is not. 
\end{itemize}

\begin{figure}[!tb]
\centering
\begin{tikzpicture}[scale=0.5, every node/.style={transform shape}]

 \node[small dot] at (-2,6) (TU2) {};
\node[small dot] at (-6,6) (TU1) {};

\node[small dot] at (2,6) (TU3) {};
\node[small dot] at (6,6) (TU4) {};

\node[small dot] at (0,3) (TM) {};

\node[small dot] at (-6,0) (TD1) {};
\node[small dot] at (-2,0) (TD2) {};

\node[small dot] at (2,0) (TD3) {};
  \node[small dot] at (6,0) (TD4) {};

 \path[draw,thick,color=black,-]
 
 (TU3) edge node {} (TM)
    (TU4) edge node {} (TM)
    (TU3) edge node {} (TU4)

    (TU1) edge node {} (TM)
    (TU2) edge node {} (TM)
    (TU1) edge node {} (TU2)
    
    (TM) edge node {} (TD1)
    (TM) edge node {} (TD2)
    (TD1) edge node {} (TD2)
    
      (TM) edge node {} (TD3)
    (TM) edge node {} (TD4)
     (TD3) edge node {} (TD4)
    ;
    
\end{tikzpicture}
\caption{The single intersection $n$-triangle network with $n=4$.}
\label{fig: unfesible KCN}
\end{figure}
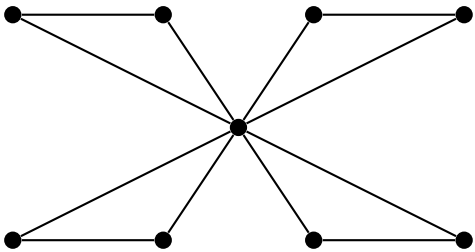

A simple instance where this holds is the \textbf{single intersection $n$-triangle} network, which consists of a central node $c$ together with $n$ triangles that include $c$ and do not share any other node. In the $1$-NIB-method, the time complexity of updating (and inference) is $\bigo \left( |X|^3 \right)$ (independently of $n$), while that of the $1$-KCN-method is at least $\bigo \left( |X|^{2(n-1)+1} \right)$, which becomes unfeasible for sufficiently large $n$. We show the single intersection $n$-triangle network with $n=4$ in Figure \ref{fig: unfesible KCN}. Recalling the inefficiencies of the KCN-method we pointed out in  Section \ref{sec: kcn-method vs bp}, one can find loopless graphs where the transition from unfeasible KCN to feasible NIB takes place. We will return to this in Section \ref{sec: nib vs bp}, where we mimic Section \ref{sec: kcn-method vs bp} but for the NIB-method. 

Even if both methods are feasible, there are instances where the suboptimal use of computational resources in the KCN-method allows us to go from innacturate KCN results to accurate NIB results without spending more computational resources:

\begin{itemize}
 \item Provided $r_0$ is a sharp loop bound and $r_0' < r_0$, there are networks where the time complexity of the $r_0$-NIB-method is bounded by that of the $r_0'$-KCN-method.
\end{itemize}
 
 An instance where this happens with $r_0=3$ and $r_0'=1$ can be found in Figure \ref{fig: triangle-square graph}, with the same holding for any network with at least two \textbf{triangle-square} pairs (see Section \ref{sec: numerics} for a definition) \footnote{Numerically, we observe that the the $1$-KCN-method achieves an error percentage $\sim 10^{-3}$ in the partition function, while the $3$-NIB-method is exact.}. While several other instances can be constructed using the intuition behind Figure \ref{fig: triangle-square graph}, characterizing all of them is not easy since the neighborhoods in each method are not related in a simple way when they use a different loop bounds.

\section{$r$-unbounded loops: The general NIB-method}
\label{sec: unbounded nib}

In this section, we do not assume some loop bound $r$ is fulfilled and define the NIB-method in general.

\subsection{Message-passing scheme}

    Since the loop bound may not be fulfilled, one cannot (see Appendix \ref{sec: definition equi classes}) define the equivalence classes on which the bounded case relies. Because of this, we have to use the neighborhoods intersections \eqref{eq: neigh inter}
    directly in order to define the algorithm. We will also have to be careful with the intersections between sets in this family, since they may no longer fulfill the simple relations in Claim \ref{claim: intersection property} and \eqref{eq: intersection equi classes}.
    
    The \textbf{NIB-method}, or \textbf{$r$-NIB-method} to be specific, is defined via a family of one-dimensional messages that sends, to each intersection of primary neighborhoods $N_{i \cap j}$, (part of) the information in $N_{k \cap q}$, the intersection of the primary neighborhoods of its constituent nodes $k \in N_{i \cap j}$ with other nodes in their primary neighborhoods $q \in N_{k}$:
     \begin{equation*}
    \begin{split}
        &\{m^{(t)}_{k \cap q \to i \cap j}\}_{i \in V, j \in N_i \smin \{i\}, k \in N_{i \cap j}, q \in  N_k,t \geq 0}, \\
    &m^{(t)}_{k \cap q \to i \cap j}: X \to \mathbb R_{\geq 0},
    \end{split}
\end{equation*}
with $m^{(t)}_{k \cap q \to i \cap j}$ depending on variable $x_k$. We only send \textbf{part} of the information to avoid unnecessary redundancies. This will become clear in the next paragraph.

The messages are uniformly initialized and, in order for the message updates to avoid redundancies as much as possible, we need a couple of definitions. First, we consider a set $\pp_{i \cap j}$, which we define recursively as follows:
\begin{itemize}
    \item we initialize it by including all the functions within $N_{i \cap j}$;
    \item at each following step, we pick some $k \in N_{i \cap j}$ and some $q \in N_k$ and we incorporate the functions within $N_{k \cap q}$ to $\pp_{i \cap j}$ \footnote{The choice of the ordering in the definition of each $\pp_{i \cap j}$ can affect the performance of the NIB-method. Good orderings can be found by inspection.}.
\end{itemize}

Given this set, we can define a map
\begin{equation*}
    \overline{\pp_{i \cap j}}: \{N_{k \cap q}\}_{k \in N_{i \cap j}, q \in N_k} \to 2^\me,
\end{equation*}
where $2^\me$ is the power set of $\me$,
such that, for each pair $k,q \in \mv$ that appears in the construction of $\pp_{i \cap j}$, $\overline{\pp_{i \cap j}}(N_{k \cap q})$ stands for $\pp_{i \cap j}$ at the step right before the pair $k,q \in \mv$ is selected.

Given these definitions, the messages are updated, for $t \geq 0$, according to the following equation:
\begin{widetext}
\begin{equation*}
\begin{split}
m_{k \cap q \to i \cap j}^{(t+1)} (x_{k}) \propto \tr_{\setminus x_{k}}  \biggl( S_{k \cap q \smin \overline{\pp_{i \cap j}}(N_{k \cap q})} \prod_{p_1 \in N_{k \cap q} \smin \{k\} } \prod_{p_2 \in N_{p_{1}} } m_{p_1 \cap p_2 \to k \cap q}^{(t)} \biggr),
\end{split}
\end{equation*}
\end{widetext}
where we have omitted a normalization constant that ensures the messages are probability distributions, and $S_{k \cap q \smin \overline{\pp_{i \cap j}}(N_{k \cap q})}$ stands for the product of the factors within $N_{k \cap j}$ which are not in $\overline{\pp_{i \cap j}}(N_{k \cap q})$.

If we denote by 
\begin{equation*}
    \{m_{k \cap q \to i \cap j}\}_{i \in V, j \in N_i \smin \{i\}, k \in N_{i \cap j}, q \in  N_k}
\end{equation*}
the set of converged messages, then we can use the equations in Appendix \ref{sec: proof inference eq} to approximately infer quantities of interest like the partition function.

While we chose to introduce the bounded case separately, there is no need to make this distinction. To see this note that, if the loop bound $r_0$ is actually fulfilled, then, for all $t \geq 0$,
\begin{equation*}
    m_{k \cap q \to i \cap j}^{(t)} \equiv m_{\overline{k \cap q} \to k}^{(t)}
\end{equation*}
provided $\overline{k \cap q}  \neq \overline{i \cap j}$ and the pair $k,q \in \mv$ is the first representative of the equivalence class $\overline{k \cap q}$ that is selected in the definition of $\pp_{i\cap j}$, while $m_{k \cap q \to i \cap j}^{(t)}$ remains uniform, and hence does not affect the results, otherwise.
 
 \subsection{Comparison to the KCN-method}
 
For the same loop bound, the time complexity of the unbounded NIB-method will always be bounded by that of the KCN-method. We can even profit from the KCN inefficiencies to go to higher loop bounds which, while still not being fulfilled, improve on the accuracy of the KCN-method:

\begin{itemize}
    \item If $r_0$ is a sharp loop bound and $r_0''<r_0' < r_0$, then there are networks where the $r_0'$-NIB-method is more accurate than the $r_0''$-KCN-method without increasing its complexity.
\end{itemize}
 
An instance of this would be the $0$-KCN- and $1$-NIB-method for the network in Figure \ref{fig: triangle-square graph}. In fact, 
this can even happen for $r_0''=r_0'<r_0$, as we have numerically observed for the same network, where $2$-NIB-method is (slightly) more accurate than the $2$-KCN-method.

The target network for which the KCN-method method was tailored and where it is expected to give accurate results are the so-called \textbf{locally dense and globally sparse} networks \cite{kirkley2021belief}. Since this notion was not rigorously defined before, we introduce the following definition:
\begin{itemize}
    \item Given a couple of integers $r_0,r_1 \geq 0$ with $r_0 \ll r_1$, a network is \textbf{$(r_0,r_1)$ locally dense and globally sparse} (\textbf{LDGS}) provided
    \begin{equation*}
N_i^{(r_0)} = N_i^{(r_0+r_1)} \subsetneq \mg        
    \end{equation*}
 for all $i \in \mv$.
\end{itemize}
That is, the set of loops is divided into two categories, with the length of those in one category being quite small w.r.t. those in the other.
In this scenario, for all $i\in \mv$ and $j \in N_i \smin\{i\}$, there exists some $\mv_{i\smin j} \subseteq \mv$ such that $N_{i \smin j} = \cup_{v \in \mv_{i\smin j}} N_{i \cap v}$, where $N_{i \cap v_1} \cap N_{i \cap v_2} = \{i \}$ for all $v_1,v_2 \in \mv_{i\smin j}$.
As a result, we have that:

\begin{claim}[Improvement for LDGS networks]
\label{claim: ldgs}
If a network is $(r_0,r_1)$ locally dense and globally sparse, then the $r_0$-NIB-method has a smaller time complexity and achieves the same accuracy as the $r_0$-KCN-method. 
\end{claim}

The proof of Claim \ref{claim: ldgs} is analogous to that of Claim \ref{claim: NIB vs kcn} \footnote{In fact, a decomposition analogous to \eqref{eq: nib decomposition of kcn} holds for locally dense and globally sparse networks as well.}. Intuitively, the NIB-method is also optimal for LDGS networks. However, extending Claim \ref{claim: NIB optimal} to this context is not analytically possible.
An instance network which is $(r_0,n)$ locally dense and globally sparse with $r_0=1$, and for which we have numerically confirmed the agreement between the methods, is the periodic $n$-triangle chain (see Figure \ref{fig: triangle cycle}) with sufficiently large $n$. We give more details in Section \ref{sec: numerics}. 

\begin{figure}[!tb]
\begin{subfigure}{0.5\textwidth}
 \centering
\begin{tikzpicture}[scale=0.5, every node/.style={transform shape}]

\node[small dot] at (0,6) (TUU) {};

\node[small dot] at (4,3) (TU4) {};
\node[small dot] at (-4,3) (TU1) {};
\node[small dot] at (2,3) (TU3) {};
\node[small dot] at (-2,3) (TU2) {};

\node[small dot] at (-4,0) (TD1) {};
\node[small dot] at (4,0) (TD3) {};
\node[small dot 2, label={[anchor=north,above=1mm, thick, font=\fontsize{18}{18}\selectfont, thick]90:\textbf{$i$}} ] at (0,0) (TD2) {};

\node[small dot 1,  label={[anchor=south,below=5mm, thick, font=\fontsize{18}{18}\selectfont, thick]90:\textbf{$j$}} ] at (0,-3) (TDD) {};

 \path[draw,thick,color=black,-]
 
 (TUU) edge node {} (TU2)
 (TUU) edge node {} (TU3)
 
 (TU1) edge node {} (TU2)
 (TU3) edge node {} (TU4)
 
 (TU1) edge node {} (TD1)
 (TU4) edge node {} (TD3)
 
  (TD2) edge node {} (TU2)
   (TD2) edge node {} (TU3)
 
 (TD1) edge node {} (TD2)
 (TD3) edge node {} (TD2)
 
  (TD2) edge node {} (TDD)
    ;
    
\end{tikzpicture}
  \caption{}
  \label{fig: extra kcn}
\end{subfigure}%
\\
\begin{subfigure}{0.5\textwidth}
   \centering
\begin{tikzpicture}[scale=0.5, every node/.style={transform shape}]

\node[small dot] at (0,6) (TUU) {};

\node[small dot 2] at (4,3) (TU4) {};
\node[small dot 2] at (-4,3) (TU1) {};
\node[small dot 2] at (2,3) (TU3) {};
\node[small dot 2] at (-2,3) (TU2) {};

\node[small dot 2] at (0,0) (TD) {};

\node[small dot] at (0,-3) (TDD) {};

\path[draw,thick,color=blue,-]
 
 (TU1) edge node {} (TU2)
 (TU2) edge node {} (TU3)
 (TU3) edge node {} (TU4)
 
 (TU1) edge node {} (TD)
 (TU2) edge node {} (TD)
 (TU3) edge node {} (TD)
 (TU4) edge node {} (TD)
    ;

 \path[draw,thick,color=black,-]
 
 (TUU) edge node {} (TU2)
 (TUU) edge node {} (TU3)
 
  (TD) edge node {} (TDD)
    ;
    
\end{tikzpicture}
  \caption{}
   \label{fig: kcn not reproducible}
\end{subfigure}%
\caption{(a) The extra computational power that the KCN-method deploys may be partially used to compute some correlations better provided some loop bound $r_0$ is not fulfilled.
For instance, taking the blue node $i$ and the red node $j$ in the figure, the correlations computed within $N_{i \smin j}$ in the $2$-KCN-method are not computed in the $2$-NIB-method. (b) The KCN-method is not always reproducible via the NIB-method. In this example, the blue region is used in the $1$-KCN-method and (in contrast with the bounded case) no equivalent set of regions is being used in any instance of the NIB-method.}
\label{fig: KCN not INT}
\end{figure}
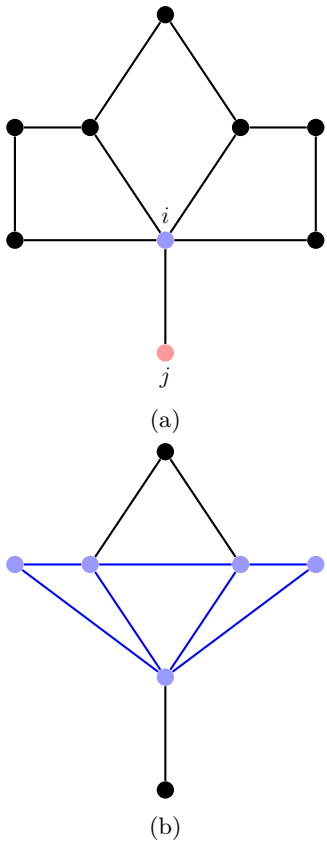

As a last comparison point, recall that, if some loop bound $r_0 \geq 0$ is fulfilled, the extra time complexity (compared to the $r_0$-NIB-method) in the $r_0$-KCN-method is useless. This also applies to any $(r_0,r_1)$ locally dense and globally sparse network. Nonetheless, this does not apply to all networks, with part of extra time complexity in the $r_0$-KCN-method sometimes allowing us to compute correlations between variables more accurately than in the $r_0$-NIB-method. An instance network where this occurs with $r_0=2$ can be found in Figure \ref{fig: extra kcn}.

\section{Computational resources in the  NIB-method}
\label{sec: nib vs bp}

As we discussed in Section \ref{sec: kcn-method vs bp}, the inefficiencies in the KCN-method become clear by comparison to network BP. In contrast, these inefficiencies are almost suppressed in the NIB-method:

\begin{claim}[$0$-NIB vs. network BP]
\label{claim: NIB vs bp}
The bounded $0$-NIB-method is equivalent to standard BP. The unbounded $0$-NIB-method provides the same accuracy as standard BP, with the complexity of its inference equations increasing by (at most) a factor of $|X|$.
  \end{claim}
  
   We prove Claim \ref{claim: NIB vs bp} in Appendix \ref{sec: other inference}.
  The statement ensures that the (bounded or unbounded) $0$-NIB-method is essentially as applicable as network BP, in stark contrast to the $0$-KCN-method. More specifically, the difference between network BP and the $0$-NIB-method does not grow with the network, while the one with the $0$-KCN-method grows exponentially with the network density.

\section{The complexity-accuracy trade-off: Interpolating between the KCN and NIB methods}
\label{sec: interpolation}

The boundary between the KCN and NIB methods is soft, making it possible to interpolate between the two of them to get new message-passing schemes. As we show in Figure \ref{fig: kcn not reproducible}, not all instances of the the KCN-method are reproducible via the NIB-method. Hence, interpolating between the methods becomes interesting in order to explore the complexity-accuracy trade-off.

By Claim \ref{claim: NIB vs kcn}, the interesting cases do not include networks where the loop bound is fulfilled. Moreover, by Claim \ref{claim: ldgs}, they also do not include locally dense and globally sparse networks. Lastly, they do not include any interpolation that involves the $0$-KCN-method. In any of these non-interesting cases we will end up with a method that provides the same accuracy as the NIB-method for a larger computational cost.

As argued in the previous paragraph, let us take a network $\mg$ and an unsatisfied loop bound $r_0 \geq 1$ such that the $r_0$-KCN-method is not reproducible via the $r_0$-NIB-method. By interpolation, and ordering them in non-decreasing complexity and (expected) non-decreasing accuracy, we obtain the following methods:

\begin{itemize}
    \item the NIB-method, where the intersection neighborhoods \eqref{eq: neigh inter} are used for both the update steps and inference;
    \item the \textbf{NIB-DIFF-method}, where the update steps use the intersection neighborhoods and the difference neighborhoods \eqref{eq: diff neigh} are used for inference; 
    \item the \textbf{KCN-NIB-method}, where the update steps use the difference neighborhoods and the intersection neighborhoods are used for inference;
    \item the \textbf{NIB-KCN-method}, where the update steps use the intersection neighborhoods and the primary neighborhoods \eqref{eq: prim neigh} are used for inference;
    \item the KCN-method, where the update steps use the difference neighborhoods and the primary neighborhoods are used for inference.
\end{itemize}

The KCN- and NIB- methods have already been described in detail before. Moreover, the NIB-KCN-method method is straightforward to envision from these two. We give some more details on the inference equations for both the NIB-DIFF and KCN-NIB methods in Appendix \ref{sec: inference interpolation case}.

\section{Numerical experiments}
\label{sec: numerics}

\begin{figure}
\begin{subfigure}{0.5\textwidth}
 \centering
\begin{tikzpicture}[scale=0.5, every node/.style={transform shape}]

\node[small dot] at (0,9) (TUU1) {};

\node[small dot] at (-2,6) (TU1) {};
\node[small dot] at (2,6) (TU2) {};

\node[small dot] at (-4,3) (TM1) {};
\node[small dot] at (4,3) (TM2) {};

\node[small dot] at (-2,0) (TD1) {};
\node[small dot] at (2,0) (TD2) {};

\node[small dot] at (0,-3) (TDD1) {};

 \path[draw,thick,color=black,-]
 
 (TUU1) edge node {} (TU1)
 (TUU1) edge node {} (TU2)
 (TU1) edge node {} (TU2)
 
 (TU1) edge node {} (TM1)
 (TM1) edge node {} (TD1)
 (TU1) edge node {} (TD1)
 
 (TU2) edge node {} (TM2)
 (TM2) edge node {} (TD2)
 (TU2) edge node {} (TD2)
 
  (TD1) edge node {} (TD2)
 (TDD1) edge node {} (TD2)
 (TDD1) edge node {} (TD1)
 
    ;
    
\end{tikzpicture}
  \caption{}
  \label{fig: triangle cycle}
\end{subfigure}%
\\
\begin{subfigure}{0.5\textwidth}
  \centering
  \begin{tikzpicture}[scale=0.5, every node/.style={transform shape}]

\node[small dot] at (-2,9) (TUU2) {};
\node[small dot] at (-6,9) (TUU1) {};

 \node[small dot] at (-2,6) (TU2) {};
\node[small dot] at (-6,6) (TU1) {};

\node[small dot] at (2,9) (TUU3) {};
\node[small dot] at (6,9) (TUU4) {};

\node[small dot] at (2,6) (TU3) {};
\node[small dot] at (6,6) (TU4) {};

\node[small dot] at (0,3) (TM) {};

\node[small dot] at (-6,0) (TD1) {};
\node[small dot] at (-2,0) (TD2) {};

\node[small dot] at (-6,-3) (TDD1) {};
\node[small dot] at (-2,-3) (TDD2) {};

\node[small dot] at (2,0) (TD3) {};
  \node[small dot] at (6,0) (TD4) {};
  
  \node[small dot] at (2,-3) (TDD3) {};
  \node[small dot] at (6,-3) (TDD4) {};

 \path[draw,thick,color=black,-]
 
 (TU3) edge node {} (TM)
    (TU4) edge node {} (TM)
    (TU3) edge node {} (TU4)

    (TU1) edge node {} (TM)
    (TU2) edge node {} (TM)
    (TU1) edge node {} (TU2)
    
    (TM) edge node {} (TD1)
    (TM) edge node {} (TD2)
    (TD1) edge node {} (TD2)
    
      (TM) edge node {} (TD3)
    (TM) edge node {} (TD4)
    (TD3) edge node  {} (TD4)
    
    (TU1) edge node  {} (TUU1)
    (TUU2) edge node  {} (TUU1)
    (TU2) edge node  {} (TUU2)
    
     (TU3) edge node  {} (TUU3)
    (TUU4) edge node  {} (TUU3)
    (TU4) edge node  {} (TUU4)
    
    (TDD3) edge node  {} (TDD4)
    (TD4) edge node  {} (TDD4)
    (TD3) edge node  {} (TDD3)
    
    (TDD1) edge node  {} (TDD2)
    (TD2) edge node  {} (TDD2)
    (TD1) edge node  {} (TDD1)
    ;
    
\end{tikzpicture}
  \caption{}
  \label{fig: triangle-square graph}
\end{subfigure}%
\caption{(a) The periodic $n$-triangle chain with $n=4$. (b) The single intersection $n$-triangle-square with $n=4$.}
\end{figure}
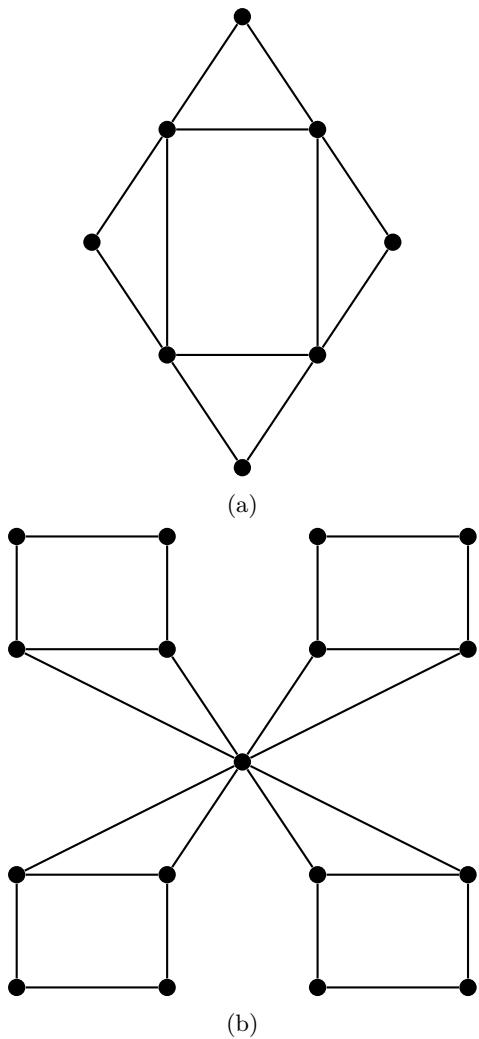

To test the performance of the NIB-method, we compare it against the KCN-method in two artificial networks with $|X|=2$:
\begin{itemize}
    \item The \textbf{periodic $n$-triangle chain} consists on $n$ triangle networks $\{t_1,\dots,t_n\}$ such that each pair $t_i,t_{i+1}$ share a single node for $i=1,\dots,n-1$ and $t_n$ and $t_1$ also share a single node. Figure \ref{fig: triangle cycle} shows the network for $n=4$.
    \item The \textbf{single intersection $n$-triangle-square} consists of $n$ pairs of triangles and squares such that each triangle is associated to a square, with which it shares two nodes, and all triangles share the node that they do not share with their associated squares. Figure \ref{fig: triangle-square graph} shows the network for $n=4$.
\end{itemize}

\begin{figure}[!tb]
\centering
    \includegraphics[scale=.85, angle=0]{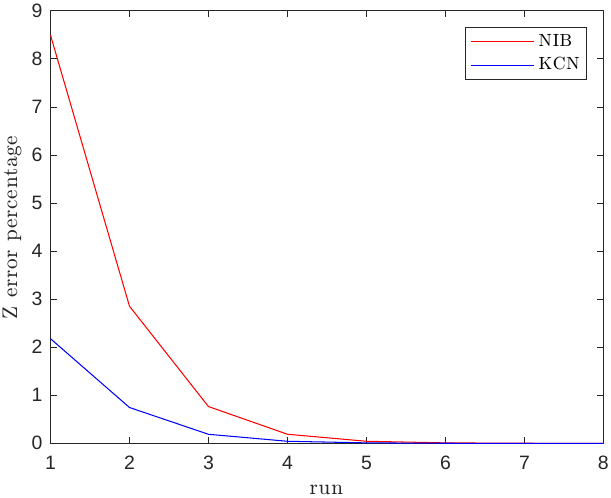}
        \caption{Variation of the partition function $Z$ mean error percentage with the number of update steps in the message-passing scheme. We have used $10^5$ random initializations of the periodic $30$-triangle chain. We include, in red, the results for $1$-NIB-method and, in blue, those for the $1$-KCN-method.}
    \label{fig: triangle cycle results}
\end{figure}

We have run $10^5$ random instances of each network, sampling the values of the factors uniformly in the interval $(0,1)$, and we have computed the mean percentage of error in the partition function using (a) the $1$-KCN and $1$-NIB methods for the periodic $n$-triangle chain, and (b) the $2$-KCN and $2$-NIB methods for the single intersection $n$-triangle-square. In both cases, we have found that, for different values of $n$ between $n=3$ and $n=30$, and for a lower computational budget, the NIB-method achieves the same computational accuracy as its KCN counterpart. In fact, for the single intersection $n$-triangle-square, and for small values of $n$, the NIB-method performs even slightly better than the KCN-method, with the improvement vanishing as the network becomes larger. Figure \ref{fig: triangle cycle results} shows, for the periodic $30$-triangle chain network, the variation of the partition function's mean percentage of error with the number of update steps.

We have also compared the KCN and 
NIB methods using the \textbf{Ising model} on these networks, that is, we have used the same networks and we have taken, for each $e=(i,j) \in \me$, $f_{i,j}(x_i,x_j) \equiv \text{exp} \left(-x_i x_j / T\right)$ instead of the random values from before, with $T>0$ playing the role of the \textbf{temperature}. We have again observed an agreement between the NIB and KCN methods.
 
 \section{Conclusion}
 \label{sec: conclusion}
 
 We have introduced here the NIB-method and compared it to the state-of-the-art generalized belief propagation algorithm, the KCN-method. We can summarize the advantages of the NIB-method as follows:
 
 \begin{itemize}
     \item If the loop bound is fulfilled, then:
 \begin{itemize}
     \item for almost every network, the NIB-method is faster than the KCN-method without loosing accuracy;
     \item no exact generalized belief propagation algorithm is faster than the NIB-method;
     \item the NIB-method can be exactly applied to a larger family of networks than the KCN-method.
 \end{itemize}
 
  \item If the loop bound is not fulfilled, then:
  \begin{itemize}
      \item the NIB-method is more efficient than the KCN-method, only using computational resources in order to compute correlations more precisely (hence, avoiding a second approximation layer when not strictly needed);
      \item the NIB-method provides, for a smaller time complexity, the same accuracy as the KCN-method for the family of networks where the KCN-method is expected to give accurate results, namely locally dense and globally sparse networks.
  \end{itemize}
   \end{itemize}
 
 Regarding further directions, it would be interesting to assess the performance of the NIB-method beyond the artificial networks that we have considered here, specially in the context of decoding, where time complexity is essential. We will address the extension of the NIB-method to other contexts in future work. In particular, its extension to other applications in network theory like percolation and the computation of matrix spectra \cite{cantwell2019message}, as well as its adaptation to more general graphical models. Regarding the later, the case of tensor networks is particularly interesting, since the use of message-passing schemes as inference tools has gained attention recently \cite{alkabetz2021tensor,guo2023block,wang2024tensor,beguvsic2024fast,hack2024belief}. Although we have been concerned with the KCN neighborhoods here, it would be interesting to apply our proposal using a network version of the WZPZ-method \cite{wang2024tensor,hack2024belief}, where, essentially, neighborhoods larger than those in the KCN-method are used and, crucially for us, the inefficiencies in the KCN-method remain \footnote{See \cite{hack2024belief} for a discussion regarding the relation between the KCN and WZPZ neighborhoods.}.  

\bibliography{main}

\clearpage

\begin{appendix}

\section{Inference in network BP}
\label{sec: inference network BP}

Given a tree network $\mg$, the following inference equations are exact:
\begin{itemize}
    \item For the marginals $p_i(\cdot)$ with $i \in \mv$, we have that
     \begin{equation*}
        p_i(x_i) \propto
\prod_{k \in \nn_i } m_{k \to i}^{(\text{BP})}(x_i),
    \end{equation*}
    where we omit a normalization constant that ensures $p_i$ is a probability distribution.
    \item For the internal energy $U$, we take the logarithm of the factors in $\mg$ as energy terms and we get
      \begin{equation*}
    U = - \sum_{(i,j) \in \me} \tr \left( \log \left( f_{i,j} \right) p_{i,j} \right),
\end{equation*}
where 
\begin{equation*}
\begin{split}
    p_{i,j}(x_i,x_j) &\equiv \frac{1}{Z_{i,j}} f_{i,j} (x_i,x_j) \prod_{k \in \nn_i \smin \{j\}} m_{k \to i}^{(\text{BP})} (x_i)  \times \\
    &\prod_{q \in \nn_j \smin \{i\}} m_{q \to j}^{(\text{BP})} (x_j),
    \end{split}
\end{equation*}
and $Z_{i,j}$ a normalization constant that ensures $p_{i,j}$ is a probability distribution.
    \item For the Shannon entropy $S$, we have that
     \begin{equation}
     \label{eq: S for net BP}
    \begin{split}
        S &= - \sum_{(i,j) \in \me} \tr \left( p_{i,j} \log p_{i,j} \right) - \sum_{i \in \mv} \left( 1-|\nn_i| \right) \tr \left( p_i \log p_i \right).
        \end{split}
    \end{equation}
    
    \item For the partition function $Z$, we have that
       \begin{equation}
    \label{eq: Z BP}
        Z = \frac{\prod_{(i,j) \in \me} \tr \left(f_{i,j} \prod_{k \in \nn_i \smin \{j\}} m_{k \to i}^{(\text{BP})} \prod_{q \in \nn_j \smin \{i\}} m_{q \to j}^{(\text{BP})} \right)}{\prod_{i \in \mv} \tr \left( \prod_{j \in \mathcal{NN}_i} m_{j \to i}^{(\text{BP})} \right)^{|\mathcal{NN}_i|-1}}.
    \end{equation}
\end{itemize}

As discussed in Section \ref{sec: network bp}, we do not use here the typical simplification of standard BP on networks. Hence, we should derive its inference equations for completeness. For brevity, and since the other equations are not more involved to derive than this one, we only do so for the partition function $Z$:

Given a tree network, \eqref{eq: Z BP} holds since, taking $n_{i \to j}^{(\text{BP})} \equiv m_{i \to j}^{(\text{BP})} / M_{i \to j}^{(\text{BP})}$, with $M_{i \to j}^{(\text{BP})}$ being the product of the normalization constants that have been accumulated in $m_{i \to j}^{(\text{BP})}$ throughout the update process, we have:
\begin{equation*}
\begin{split}
     &\frac{\prod_{(i,j) \in \me} \tr \left(f_{i,j} \prod_{k \in \nn_i \smin \{j\}} m_{k \to i}^{(\text{BP})} \prod_{q \in \nn_j \smin \{i\}} m_{q \to j}^{(\text{BP})} \right)}{\prod_{i \in \mv} \tr \left( \prod_{j \in \mathcal{NN}_i} m_{j \to i}^{(\text{BP})} \right)^{|\nn_i|-1}} \\
     &= \frac{\prod_{(i,j) \in \me} \tr \left(f_{i,j} \prod_{k \in \nn_i \smin \{j\}} n_{k \to i}^{(\text{BP})} \prod_{q \in \nn_j \smin \{i\}} n_{q \to j}^{(\text{BP})} \right)}{\prod_{i \in \mv} \tr \left( \prod_{j \in \mathcal{NN}_i} n_{j \to i}^{(\text{BP})} \right)^{|\nn_i|-1}} \\
     &= \frac{\prod_{(i,j) \in \me} Z}{\prod_{i \in \mv} Z^{|\nn_i|-1}} =  Z^{|\me| - \left( \sum_{i \in \mv} \left( |\nn_i| -1  \right) \right)}\\
     &= Z^{|\mv| -1 -\left( 2 \left(|\mv|-1\right) - |\mv| \right)} = Z,
     \end{split}
\end{equation*}
where we have cancelled the normalization constants $M_{i \to j}^{(\text{BP})}$ in the first equality since they appear exactly $|\nn_i|-1$ times in the numerator and only once in the denominator, and we have used \cite[Theorem 9.1]{wilson1979introduction} in the second to last.

When facing a network with loops, we can use the previous equations for approximate inference.

It is important to note that the essence of these equations is the substitution of the probability distribution associated to a network by another probability distribution composed of independent terms associated to some of its subnetworks. More specifically, network BP is based on the following exact decomposition for tree networks:
\begin{widetext}
\begin{equation}
\label{eq: prob decomp net BP}
\begin{split}
    p(x_1,\dots,x_{|\mv|}) \propto &\prod_{(i,j) \in \me} \left(f_{i,j} (x_i,x_j) \prod_{k \in \nn_i \smin \{j\}} m_{k \to i}^{(\text{BP})} (x_i) \prod_{q \in \nn_j \smin \{i\}} m_{q \to j}^{(\text{BP})} (x_j) \right)\\
    &\times \prod_{i \in \mv} \left( \prod_{j \in \mathcal{NN}_i} m_{j \to i}^{(\text{BP})} (x_i) \right)^{1-|\mathcal{NN}_i|}.
    \end{split}
\end{equation}
\end{widetext}
This decomposition is useful to obtain some of the inference equations from the rest. For instance, under the loopless assumption, we can use \eqref{eq: prob decomp net BP} to get another exact equation for the internal energy
\begin{widetext}
\begin{equation}
\label{eq: U for net BP II}
\begin{split}
    U = &-\sum_{(i,j) \in \me} \tr \left( p_{i,j} \left( \log \left( f_{i,j} \right) + \sum_{k \in \nn_i \smin \{j\}} \log \left(m_{k \to i}^{(\text{BP})} \right) + \sum_{q \in \nn_j \smin \{i\}} \log \left( m_{q \to j}^{(\text{BP})} \right) \right) \right) \\
    &-\sum_{i \in \mv} \left( 1- |\nn_i| \right) \tr \left( p_i \sum_{j \in \mathcal{NN}_i} \log \left( m_{j \to i}^{(\text{BP})} \right)  \right),
    \end{split}
\end{equation}
\end{widetext}
which we can use together with \eqref{eq: S for net BP} and the well-known equation
\begin{equation}
\label{eq: basic stat mech}
    Z = e^{S-U}
\end{equation}
to obtain \eqref{eq: Z BP}.

Importantly, if the loopless assumption does not hold, all the inference equations are based on the same assumption, namely that the decomposition \eqref{eq: prob decomp net BP} holds, and we can think of them as being determined once the decomposition is fixed. This will be relevant when discussing inference in the KCN-method in Appendix \ref{sec: kcn inference}.

\section{Inference in the KCN-method}
\label{sec: kcn inference}

The inference equations for the KCN-method were already provided when it was introduced \cite{kirkley2021belief} (see for instance \cite[Eq. 11]{kirkley2021belief} for the marginals, \cite[Eq. 14]{kirkley2021belief} for the internal energy, and \cite[Eq. 22]{kirkley2021belief} for the Shannon entropy). We devote this section to making a couple of comments on them. 

Our first comment concerns the inference of the partition function. Despite it being the central quantity of interest \cite{kirkley2021belief}, no direct equation in terms of messages is derived for it. More specifically, in contrast to their equations for the internal energy of the Shannon entropy, the partition function is derived via \eqref{eq: basic stat mech}. This is interesting when contrasted with the discussion regarding network decomposition in Appendix \ref{sec: inference network BP}. In particular, \cite{kirkley2021belief} deploys one sort of network decomposition when dealing with the internal energy and a different one when dealing with the Shannon entropy. However, we can use the composition from the Shannon entropy (the one for the internal energy is not general enough) to obtain an inference equation for the partition function:
 \begin{widetext}
 \begin{equation}
 \label{eq: partition kcn}
     Z = \frac{\prod_{((i,j)) \in \mg} \tr \left(S_{i \cap j} m_{j \to i}^{(\text{KCN})} \prod_{k \in N_{i \cap j} \smin \{j\}} m_{k \to j}^{(\text{KCN})} \right)^\frac{1}{\binom{|S_{i \cap j}|}{2}}}{\prod_{(i,j) \in \me} \tr \left(S_{i \cap j} m_{j \to i}^{(\text{KCN})} \prod_{k \in N_{i \cap j} \smin \{j\}} m_{k \to j}^{(\text{KCN})} \right)^{-W_{i,j}} \prod_{i \in \mv} \tr \left(S_i  \prod_{j \in N_i \smin \{i\}} m_{j \to i}^{(\text{KCN})} \right)^{-C_i}},
 \end{equation}
 \end{widetext}
 where $S_i$ stands for the functions within $N_i$, $((i,j))$ stands for the pairs $i,j \in \mg$ such that $i \in N_j$ and, hence, $j \in N_i$, $|S_{i \cap j}|$ stands for the number of variables within $S_{i \cap j}$,  and 
 \begin{equation}
 \label{eq: coefficients w and c}
    \begin{split}
        W_{i,j} &\equiv 1 - \sum_{((\ell,k)) \in \mg} \frac{1}{\binom {|S_{\ell \cap k}|}{2}} \chi_{\{(i,j) \in N_{\ell,k}\}},\\
        C_i &\equiv 1- \sum_{j \in N_i} \left( \frac{1}{|S_{i \cap j}|-1} \right) - \sum_{j \in \mathcal{NN}_i} W_{i,j}.
    \end{split}
\end{equation}
Note that, if $r=0$, then \eqref{eq: partition kcn} holds with $W_{i,j} = 0$ for all $(i,j) \in \mg$. We will focus on this case in Appendix \ref{sec: other inference}. 
 
 Our second comment is a short clarification regarding the inference of the two-variable marginals $p_{i,j}$, which appear in the inference equation for the Shannon entropy in the KCN-method. In general \cite{mezard2009information,kirkley2021belief}, to get such a marginal for any couple of variables requires running a message-passing method (say the KCN-method) once to estimate a single variable marginal $p_i(\cdot)$ and then running it again, once for each value $x_i \in X$, in order to obtain the single variable marginal $p_j^{x_i}$ conditioned on $X_i=x_i$. This process is computationally expensive. Nonetheless, in the KCN-method we are only interested in $p_{i,j}$ for $(i,j) \in \me$ and, hence, we can estimate it via
 \begin{equation*}
 \begin{split}
     p_{i,j}(x_i,x_j) &\propto \frac{1}{2 Z_i} \tr_{\smin(x_i,x_j)} \left(S_i \prod_{k \in N_i \smin \{i\}} m_{k \to i}^{(\text{KCN})} \right) +\\
     & \frac{1}{2 Z_j} \tr_{\smin(x_i,x_j)} \left(S_j \prod_{k \in N_j \smin \{j\}} m_{k \to j}^{(\text{KCN})} \right),
 \end{split}
 \end{equation*}
 where $Z_i$ and $Z_j$ are normalization constants that ensure each of the summands is a probability distribution. Later on, we will use a similar equation for $p_{i,j}$ which is based on the intersection set $N_{i \cap j}$.

\section{Computational resources in the KCN and NIB methods}
\label{sec: other inference}

\subsection{Proof of Claim \ref{claim: kcn vs bp}}

  By definition, for all $t \geq 0$, we have the following relations:
  \begin{equation}
  \label{eq: relation messages}
      \begin{split}
          &m_{j \to i}^{(\text{KCN},t)} (x_j) \propto \prod_{k \in \mathcal{NN}_j \smin \{i\}} m_{k \to j}^{(\text{BP},t-1)} (x_j), \\
          &m_{j \to i}^{(\text{BP},t)} (x_i) \propto \sum_{x_j} f_{i,j} (x_i,x_j) m_{j \to i}^{(\text{KCN},t-1)} (x_j).
      \end{split}
  \end{equation}
  These equations imply that, if one of the methods converges, then the other one does as well. Assume, hence, that both methods have converged, respectively, to the messages
  \begin{equation*}
      \begin{split}
          &\{m^{(\text{BP})}_{i \to j}\}_{i \in \mv, j \in \mathcal{NN}_i,t \geq 0} \text{ and} \\
        &\{m^{(\text{KCN})}_{i \to j}\}_{i \in \mv, j \in \mathcal{NN}_i,t \geq 0}.        
      \end{split}
  \end{equation*}
   We conclude by showing that they infer the same quantities:

  Regarding marginals, we have that
\begin{equation*}
\begin{split}
     p_i^{(\text{KCN})}(x_i) &\propto \tr_{\smin x_i} \left( \prod_{j \in \mathcal{NN}_i} f_{i,j} m_{j \to i}^{(\text{KCN})} \right) \\
     &\propto \prod_{j \in \mathcal{NN}_i}\tr_{\smin x_i} \left( f_{i,j} m_{j \to i}^{(\text{KCN})} \right) \\
     &\propto \prod_{j \in \mathcal{NN}_i} m_{j \to i}^{(\text{BP})} (x_i) \\
     &\propto p_i^{(\text{BP})}(x_i).
\end{split}
\end{equation*}
This implies that $p_i^{(\text{KCN})} =  p_i^{(\text{BP})}$ by normalization.

Regarding internal energy, we have that
\begin{widetext}
\begin{equation*}
\begin{split}
     U^{(\text{BP})}&= -\sum_{(i,j) \in \me} \tr \left( \log f_{i,j} \frac{f_{i,j} \prod_{k \in \nn_i \smin \{j\}} m_{k \to i}^{(\text{BP})} \prod_{q \in \nn_j \smin \{i\}} m_{q \to j}^{(\text{BP})}}{\tr \left( f_{i,j} \prod_{k \in \nn_i \smin \{j\}} m_{k \to i}^{(\text{BP})} \prod_{q \in \nn_j \smin \{i\}} m_{q \to j}^{(\text{BP})} \right)} \right)  \\
     &= -\sum_{(i,j) \in \me} \tr \left( \log f_{i,j} \frac{f_{i,j} m_{i \to j}^{(\text{KCN})} m_{j \to i}^{(\text{KCN})}}{\tr \left( f_{i,j} m_{i \to j}^{(\text{KCN})} m_{j \to i}^{(\text{KCN})} \right)} \right) \\
     &= -\frac{1}{2} \sum_{i \in \mv, j \in \nn_i} \tr \left( \log f_{i,j} \frac{ \prod_{j \in \nn_i} f_{i,j} m_{j \to i}^{(\text{KCN})}}{\tr \left( \prod_{j \in \nn_i} f_{i,j} m_{j \to i}^{(\text{KCN})} \right)} \right) \\
     &= U^{(\text{KCN})},
\end{split}
\end{equation*}
\end{widetext}
where we have used \eqref{eq: relation messages} in the second equality, and the fact that the messages have converged in the third equality. 

Regarding entropy, we have that 
\begin{widetext}
\begin{equation*}
\begin{split}
     S^{(\text{KCN})}&= -\sum_{(i,j) \in \me} \tr \left( \frac{f_{i,j} m_{i \to j}^{(\text{KCN})}m_{j \to i}^{(\text{KCN})}}{\tr \left( f_{i,j} m_{i \to j}^{(\text{KCN})} m_{j \to i}^{(\text{KCN})} \right)} \log \left( \frac{f_{i,j} m_{i \to j}^{(\text{KCN})}m_{j \to i}^{(\text{KCN})}}{\tr \left( f_{i,j} m_{i \to j}^{(\text{KCN})} m_{j \to i}^{(\text{KCN})} \right)} \right) \right) \\ &-\sum_{i \in \mv} \left(1 - |\nn_i| \right) \tr \left( \frac{\tr_{\smin x_i} \left( \prod_{j \in \nn_i} f_{i,j} m_{j \to i}^{(\text{KCN})}\right)}{\tr \left( \prod_{j \in \nn_i} f_{i,j} m_{j \to i}^{(\text{KCN})} \right)} \log \left( \frac{\tr_{\smin x_i} \left(\prod_{j \in \nn_i} f_{i,j} m_{j \to i}^{(\text{KCN})}\right)}{\tr \left( \prod_{j \in \nn_i} f_{i,j} m_{j \to i}^{(\text{KCN})} \right)} \right) \right)  \\
    &= -\sum_{(i,j) \in \me} \tr \Bigg( \frac{f_{i,j} \prod_{k \in \nn_i \smin \{j\}} m_{k \to i}^{(\text{BP})} \prod_{q \in \nn_j \smin \{i\}} m_{q \to j}^{(\text{BP})}}{\tr \left( f_{i,j} \prod_{k \in \nn_i \smin \{j\}} m_{k \to i}^{(\text{BP})} \prod_{q \in \nn_j \smin \{i\}} m_{q \to j}^{(\text{BP})} \right)} \times \\
    &\log \left( \frac{f_{i,j} \prod_{k \in \nn_i \smin \{j\}} m_{k \to i}^{(\text{BP})} \prod_{q \in \nn_j \smin \{i\}} m_{q \to j}^{(\text{BP})}}{\tr \left( f_{i,j} \prod_{k \in \nn_i \smin \{j\}} m_{k \to i}^{(\text{BP})} \prod_{q \in \nn_j \smin \{i\}} m_{q \to j}^{(\text{BP})} \right)}\right) \Bigg) \\ 
    &-\sum_{i \in \mv} \left(1 - |\nn_i| \right) \tr \left( \frac{\prod_{k \in \nn_i} m_{k \to i}^{(\text{BP})}}{\tr \left( \prod_{k \in \nn_i} m_{k \to i}^{(\text{BP})} \right)} \log \left( \frac{\prod_{k \in \nn_i} m_{k \to i}^{(\text{BP})}}{\tr \left( \prod_{k \in \nn_i} m_{k \to i}^{(\text{BP})} \right)} \right) \right)  \\
     &= S^{(\text{BP})},
\end{split}
\end{equation*}
\end{widetext}
where we have used \eqref{eq: relation messages} in the second equality.

Regarding the partition function, we can proceed along the lines of the entropy equality to get $Z^{(\text{KCN})} = Z^{(\text{BP})}$, with $Z^{(\text{KCN})}$ defined as in \eqref{eq: partition kcn}.

It is important to note that we have not required the network to be a tree at any step.

\subsection{Proof of Claim \ref{claim: NIB vs bp}}

 In the bounded case, we have that
  \begin{equation*}
  m_{\overline{i \cap j} \to i}^{(\text{NIB},t)} \equiv m_{i \to j}^{(\text{BP},t)}.    
  \end{equation*}
  Moreover, the inference equations also coincide.
  
  In the unbounded case, and provided there exists some $k \in \nn_j \smin \{i\}$, we have that
  \begin{equation*}
  m_{i \cap j \to j \cap k}^{(\text{NIB},t)} \equiv m_{i \to j}^{(\text{BP},t)}.    
  \end{equation*}
  This means that network BP computes more messages than the unbounded $0$-NIB-method, since $m_{i \to j}^{(\text{BP},t)}$ has no NIB counterpart provided $\nn_j \smin \{i\} = \emptyset$.
  
  The relation among the messages implies that, if one of the methods converges, then the other one does as well. To convince ourselves of this, we simply note that the messages $m_{i \to j}^{(\text{BP},t)}$ which are not computed in the unbounded $0$-NIB-method do not affect the rest of the messages, since we have $\nn_j \smin \{i\} = \emptyset$.
  
  From the convergence assumption and the inference equations, it becomes clear that the inferred quantities are the same. Regarding time complexity, the methods only differ in the one associated to inference, where the NIB-method requires, at most, a factor of $|X|$ more time. This increase takes place in the estimation of marginals, where the complexity rises form $|X|$ to $|X|^2$. Regarding other quantities, the complexity of the partition function doubles, while that of the internal energy stays the same.
  
  \section{Properties of neighborhoods intersections}
  \label{sec: definition equi classes}
  
  \subsection{The equivalence class condition:\\ Proof of Claim \ref{claim: intersection property}}

To show the result, it suffices to argue that, given $k \in N_{i \cap j}$, we have $N_{i \cap j} \subseteq N_k$. This is the case since, if this holds, it also does for $q \in N_{i \cap j}$, and we get that $N_{i \cap j} \subseteq N_{k \cap q}$. Since $i,j \in N_{k \cap q}$, we analogously get that $N_{k \cap q} \subseteq N_{i \cap j}$ and we obtain the result $N_{i \cap j} = N_{k \cap q}$.

To conclude, hence, we show that $N_{i \cap j} \subseteq N_k$ provided $k \in N_{i \cap j}$. Note first that it, if we show the result for edges $\me \cap N_{i \cap j} \subseteq N_{k}$, then the result holds for vertices $\mv \cap N_{i \cap j} \subseteq N_{k}$ by definition.

We show in what follows that $e \in N_k$ for any edge $e \in \me \cap N_{i \cap j}$. In order to do so, let us first construct a path $p_{i-e-j}$ that goes from $i$ to $j$ through $e$: We start by taking a cycle $c_{i-e-i}$ from $i$ to itself that passes through $e$. We also consider a path $p_{j-e}$ that starts at $j$ and finishes at $e$, and we take $r_0 \in \mv$ the closest vertex to $j$ in $p_{j-e}$ such that $r_0 \in c_{i-e-i} \cap p_{j-e}$. (When referring to two elements $x,y$ belonging to a path $p$, by \textbf{distance}, and hence by \textbf{closest}, with respect to $p$ we refer to the number of edges and vertices that separate $x$ from $y$ along $p$.) We take
\begin{equation*}
    p_{i-e-j} \equiv p_{j-r_0} \cup p_{r_0-e-i}, 
\end{equation*}
where $p_{j-r_0}$ is the path within $p_{j-e}$ that joins $j$ and $r_0$, and $p_{r_0-e-i}$ is the path within $c_{i-e-i}$ that goes from $r_0$ to $i$ through $e$. By construction, $p_{i-e-j}$ is a path.

We now want to use $p_{i-e-j}$ to form a cycle around $k$ that contains $e$. To do so, note that $p_{i-e-j}$ intersects with $c_{k-j-k}$ and $c_{k-i-k}$, which are, respectively, cycles that include $k$ and $j$, and $k$ and $i$. We denote by $r_i \in \mv$ (and analogously by $r_j \in \mv$) the vertex between $e$ and $i$ ($e$ and $j$) that is closest to $e$ in $p_{i-e-j}$ such that $r_i \in c_{k-j-k} \cup c_{k-i-k}$ ($r_j \in c_{k-j-k} \cup c_{k-i-k}$), and note that $r_i \neq r_j$.

We conclude by considering two cases:
\begin{enumerate}[label=(\Roman*)]
    \item We can take $c \in \{c_{k-j-k}, c_{k-i-k}\}$ such that $r_i,r_j \in c$. Then, we take
     \begin{equation*}
    c_{k-e-k} \equiv p_{k-r_i} \cup p_{r_i-e-r_j} \cup p_{r_j-k},    
    \end{equation*}
    where $p_{k-r_i}$ and $p_{r_j-k}$ are the subpaths of $c$ that, respectively, join $r_i$ and $k$, and $r_j$ and $k$, and fulfill $p_{k-r_i} \cap p_{r_j-k} = \{k\}$, and  $p_{r_i-e-r_j}$ the subpath of $p_{i-e-j}$ that joins $r_i$ and $r_j$. By construction, 
   $c_{k-e-k}$ is a cycle around $k$ that includes $e$ and, hence, $e \in N_k$.

    \item If $r_i \in c_1$ and $r_j \in c_2$, with $c_1,c_2 \in \{c_{k-j-k}, c_{k-i-k}\}$ and $c_1 \neq c_2$, then, assuming w.l.o.g. that $r_i \in c_{k-i-k}$, we take
    \begin{equation*}
       c_{k-e-k} \equiv p_{k-s} \cup p_{s-r_i} \cup p_{r_i-e-r_j} \cup p_{r_j-k},
    \end{equation*}
    where, taking $s$ to be the closest vertex to $r_i$ in $c_{k-i-k}$ that also belongs to $c_{k-j-k}$
    (we assume w.l.o.g. $s \neq r_j$ since, otherwise, we are in case (I)),
    $p_{s-r_i}$ is the subpath of $c_{k-i-k}$ that joins $r_i$ with $s$, $p_{k-s}$ the subpath of $c_{k-j-k}$ that joins $s$ and $k$ without going through $r_j$, and $p_{r_j-k}$ is the subpath of $c_{k-j-k}$ that joins $r_j$ and $k$ without going through $s$. By construction, $c_{k-e-k}$
    is a cycle around $k$ that includes $e$ and, hence, $e \in N_k$.
\end{enumerate}

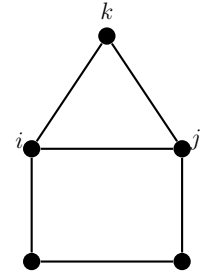
\begin{figure}[!tb]
\centering
\begin{tikzpicture}[scale=0.5, every node/.style={transform shape}]
\node[small dot, label={[anchor=north,above=1mm, thick, font=\fontsize{18}{18}\selectfont, thick]90:\textbf{$k$}} ] at (0,6) (TUU) {};

\node[small dot, label={[anchor=east,right=1mm, thick, font=\fontsize{18}{18}\selectfont, thick]90:\textbf{$j$}} ] at (2,3) (TU2) {};
\node[small dot, label={[anchor=west,left=1mm, thick, font=\fontsize{18}{18}\selectfont, thick]90:\textbf{$i$}} ] at (-2,3) (TU1) {};

\node[small dot] at (2,0) (TD2) {};
\node[small dot] at (-2,0) (TD1) {};

 \path[draw,thick,color=black,-]
 
 (TUU) edge node {} (TU1)
 (TUU) edge node {} (TU2)
 
 (TU1) edge node {} (TD1)
 (TU2) edge node {} (TD2)
 
 (TU1) edge node {} (TU2)
 (TD1) edge node {} (TD2)
    ;
    
\end{tikzpicture}
\caption{If loop bound is not satisfied, then the equivalence class condition may not hold. In this instance, taking $r_0=2$, we have that $k \in N_{i \cap j} = \mg$ while $N_k \subsetneq \mg$. Moreover, if the equivalence class condition holds for some $r_0 \geq 0$, this does not imply that $r_0$ is a loop bound. In this instance, although the condition holds for $r_0=1$, this is not a fulfilled loop bound.}
\label{fig: no unbounded equivalences}
\end{figure}

Note that the hypothesis on the loop bound is key, since the equivalence class condition does not hold provided it is not fulfilled. Moreover, the converse of the claim is also false. Figure \ref{fig: no unbounded equivalences} shows a network that serves as a counterexample for both these two statements.

\subsection{Characterization of the neighborhoods intersections}

\begin{claim}[Neighborhoods intersections characterization]
\label{claim: charact intersection}
If the loop bound $r_0 \geq 0$ is fulfilled, $i \in \mv$, and $j \in \mv \cap \left( N_i^{(r_0)} \smin \{i \} \right)$, then $N_{i \cap j}^{(r_0)} = C_{i-j} \cup P_{C_{i-j}}$, where
\begin{equation*}
\begin{split}
    &C_{i-j} \equiv \{c \subseteq \mg| c \text{ is a cycle and } i,j \in c\}\\
    &P_{C_{i-j}} \equiv \{p \subseteq \mg| p \text{ is a path and } |\mv \cap p \cap c| = 2 \\
    &\text{for some } c \in C_{i-j}\}.
    \end{split}
\end{equation*}
\end{claim}

To prove the claim, by begin by showing that
\begin{equation*}
    C_{i-j} \cup P_{C_{i-j}} \subseteq N_{i \cap j}^{(r_0)}.
\end{equation*}
 By definition, $C_{i-j} \subseteq N_{i \cap j}^{(r_0)}$, while, taking as $i_1,i_2 \in \mv$ the intersections of $p$ with some $c \in C_{i-j}$, it is clear that $p$ belongs to a cycle around $i$ ($j$) that consists of $p$ together with the segment of $c$ that joins  $i_1$ and $i_2$ through $i$ ($j$). In particular, 
$P_{C_{i-j}} \subseteq N_{i \cap j}^{(r_0)}$

We conclude by showing that $N_{i \cap j}^{(r_0)} \subseteq C_{i-j} \cup P_{C_{i-j}}$. To do so, it suffices to show that
\begin{equation*}
\me \cap N_{i \cap j}^{(r_0)} \subseteq \me \cap (C_{i-j} \cup P_{C_{i-j}}).    
\end{equation*}
Consider, hence, some $e \in \me \cap N_{i \cap j}^{(r_0)}$ and assume that there is no $c \in C_{i-j}$ such that $e \in c$. By definition, there exist some cycle $c_i$ ($c_j$) such that $i,e \in c_i$ ($j,e \in c_j$). Moreover, if $i \in c_j$, then $e \in c_j \in C_{i-j}$, a contradiction. Hence, $i \not \in c_j$ and we can consider $p_{b-i-a}$ the path contained in $c_i$ that goes through $i$ and has as endpoints $a,b$ the two nodes (one for each branch of $c_i$) in $c_i$ that are the closest to $i$ and intersect $c_j$ (they must intersect it since $e \in c_i \cap c_j$). Since $a \neq b$ (otherwise $e \not \in c_j$, a contradiction), we have that
\begin{equation*}
c_{i-j} \equiv p_{b-i-a} \cup p_{a-j-b}    
\end{equation*}
is a cycle such that $i,j \in c_{i-j}$, where $p_{a-j-b}$ is the segment of $c_j$ that joins $a$ and $b$ going through $j$. Since $e \not \in c_{i-j}$ by assumption, we have that
\begin{equation*}
 e \in p_{a-e-b} \equiv c_j \smin c_{i-j},   
\end{equation*}
 where $p_{a-e-b}$ consists of the edges in $c_j$ which are not in $c_{i-j}$ together with the endpoints of such edges. This concludes the argument since $p_{a-e-b}$ is a path such that $p_{a-e-b} \cap c_{i-j} = \{a,b\}$.

\section{Inference in the NIB-method}
\label{sec: proof inference eq}

\subsection{r-bounded loops}
\label{sec: proof inference eq BOUNDED}

Given a network $\mg$ where the loop bound is fulfilled, the following inference equations are exact:
\begin{itemize}
    \item For the marginals $p_i(\cdot)$ with $i \in \mv$, we have that
     \begin{equation}
    \label{eq: marginals intersect}
        p_i(x_i) \propto \prod_{\overline{i \cap j}} m_{\overline{i \cap j} \to i}(x_i),
    \end{equation}
    where we omit a normalization constant that ensures $p_i$ is a probability distribution.
    \item For the internal energy $U$, we take the logarithm of the functions in $\mg$ as energy terms and we get
      \begin{equation*}
    U = - \sum_{(i,j) \in \me} \frac{1}{Z_{i \cap j}} \tr \left( \log f_{i,j} S_{i \cap j} \prod_{k \in N_{i \cap j}} \prod_{\overline{k \cap q} \neq \overline{i \cap j} } m_{\overline{k \cap q} \to k} \right),
\end{equation*}
where $Z_{i \cap j}$ is a normalization constant
 \begin{equation*}
   Z_{i \cap j} \equiv \tr \left( S_{i \cap j} \prod_{k \in N_{i \cap j}} \prod_{\overline{k \cap q} \neq \overline{i \cap j} } m_{\overline{k \cap q} \to k} \right).
\end{equation*}
    \item For the Shannon entropy $S$, we have that
       \begin{equation*}
\begin{split}
    S = &- \sum_{\overline{i \cap j} \in \equi{\cap}} \sum_{x_{i \cap j}} p_{i \cap j}(x_{i \cap j}) \log p_{i \cap j}(x_{i \cap j}) \\
    &- \sum_{i \in \mg} \left( 1-\left|\{\overline{k \cap q} \in \equi{\cap} | i \in \overline{k \cap q} \}\right| \right) \sum_{x_i} p_i(x_i) \log p_i(x_i),
\end{split}
\end{equation*}
where the marginals $p_i$ are computed via \eqref{eq: marginals intersect} and
\begin{equation}
\label{eq: intersection marginal NIB}
    p_{i \cap j} (x_{i \cap j}) \propto S_{i \cap j} \prod_{k \in N_{i \cap j}} \prod_{\overline{k \cap q} \neq \overline{i \cap j} } m_{\overline{k \cap q} \to k},
\end{equation}
where we omit a normalization constant that ensures $p_{i \cap j}$ is a probability distribution.
    
    \item For the partition function $Z$, we have that
   \begin{equation}
\label{eq: intersection Z II}
Z = \frac{\prod_{\overline{i \cap j} \in \equi{\cap}} \tr \left( S_{i \cap j} \prod_{k \in N_{i \cap j}} \prod_{\overline{k \cap q} \neq \overline{i \cap j} } m_{\overline{k \cap q} \to k} \right)}{\prod_{i \in \mg} \tr \left( \prod_{\overline{i \cap j}} m_{\overline{i \cap j} \to i} \right)^{\left|\{\overline{k \cap q} \in \equi{\cap} | i \in \overline{k \cap q} \}\right|-1}}.
\end{equation}
\end{itemize}

For brevity, and since the other equations are at most as difficult as this one, we only show the inference equation for the Shannon entropy $S$:

To prove the entropy formula, it is sufficient to show that, provided the loop bound is fulfilled, we have that
\begin{equation}
\label{eq: probability kcn}
    p(x) = \prod_{\overline{i\cap j} \in \equi{\cap}} p_{i \cap j}(x_{i \cap j}) \prod_{i \in \mv} p_i(x_i)^{1-\left|\left\{\overline{k \cap q} \in \equi{\cap} | i \in \overline{k \cap q} \right\}\right|}.
\end{equation}
This relation can be proven as follows:
\begin{widetext}
\begin{equation*}
\begin{split}
    &\prod_{\overline{i\cap j} \in \equi{\cap}} p_{i \cap j}(x_{i \cap j}) \prod_{i \in \mv} p_i(x_i)^{1-\left|\left\{\overline{k \cap q} \in \equi{\cap} | i \in \overline{k \cap q} \right\}\right|} \\
    &= \prod_{\overline{i\cap j} \in \equi{\cap}} \left( \frac{1}{Z_{\overline{i \cap j}}} S_{i \cap j} (x_{i \cap j}) \prod_{k \in N_{i \cap j}} \prod_{\overline{k \cap q} \neq \overline{i \cap j} } m_{\overline{k \cap q} \to k} (x_k) \right) \prod_{i \in \mv} \left( \frac{1}{Z_{i}} \prod_{\overline{i \cap j}} m_{\overline{i \cap j} \to i}(x_i) \right)^{1-\left|\left\{\overline{k \cap q} \in \equi{\cap} | i \in \overline{k \cap q} \right\}\right|} \\
    &= \prod_{\overline{i\cap j} \in \equi{\cap}} \left( \frac{1}{Z} S_{i \cap j} (x_{i \cap j}) \prod_{k \in N_{i \cap j}} \prod_{\overline{k \cap q} \neq \overline{i \cap j} } n_{\overline{k \cap q} \to k} (x_k) \right) \prod_{i \in \mv} \left( \frac{1}{Z} \prod_{\overline{i \cap j}} n_{\overline{i \cap j} \to i}(x_i) \right)^{1-\left|\left\{\overline{k \cap q} \in \equi{\cap} | i \in \overline{k \cap q} \right\}\right|}\\
    &= \frac{1}{Z} \left( \prod_{\overline{i\cap j} \in \equi{\cap}} S_{i \cap j} (x_{i \cap j}) \right) \prod_{i \in \mv} \left(\prod_{\overline{i \cap j}} n_{\overline{i \cap j} \to i}(x_i) \right)^{\left|\{\overline{k \cap q} \in \equi{\cap} | i \in \overline{k \cap q} \}\right|-1} \prod_{i \in \mv} \left( \prod_{\overline{i \cap j}} n_{\overline{i \cap j} \to i}(x_i) \right)^{1-\left|\left\{\overline{k \cap q} \in \equi{\cap} | i \in \overline{k \cap q} \right\}\right|}\\
    &= p(x),
    \end{split}
\end{equation*}
\end{widetext}
where we have used \eqref{eq: intersection marginal NIB} and \eqref{eq: marginals intersect} in the first equality, together with $Z_{\overline{i \cap j}}$ and $Z_i$ to refer, respectively, to the missing normalization constants in them. In the second equality, we have used the definition
$n_{\overline{k \cap q} \to k} \equiv m_{\overline{k \cap q} \to k} / M_{\overline{k \cap q} \to k}$, with $M_{\overline{k \cap q} \to k}$ being the product of the normalization constants that have been accumulated in $m_{\overline{k \cap q} \to k}$ throughout the update process,
and we have
noted that the product of the missing normalization constants $M_{\overline{k \cap q} \to k}$ together with $Z_{\overline{i \cap j}}$ and $Z_i$, respectively, give us the partition function for each $\overline{i\cap j} \in \equi{\cap}$ and $i \in \mv$.
In the third equality, we have used the hypergraph analog of \cite[Theorem 9.1]{wilson1979introduction}. 

\subsection{r-unbounded loops}

Given a network $\mg$ where the loop bound is not fulfilled, we use the following approximate inference equations:
\begin{itemize}
    \item For the marginals $p_i(\cdot)$ with $i \in \mv$, and  omitting a normalization constant that ensures $p_i$ is a probability distribution, we use
    \begin{equation}
    \label{eq: approx marginals intersect}
    \begin{split}
        p_i(x_i) &\propto \prod_{j \in N_i \smin \{i\}} \tr_{\setminus x_{i}}  \biggl( S_{i \cap j \smin \overline{\ppq_{i}}(N_{i \cap j})}\\
        &\times \prod_{p_1 \in N_{i \cap j} \smin \{i\} } \prod_{ p_2 \in N_{p_1} } m_{p_1 \cap p_2 \to i \cap j} \biggr),
         \end{split}
    \end{equation}
    where $S_{i \cap j \smin \overline{\ppq_{i}}(N_{i \cap j})}$ stands for the product of the factors which are within $N_{i \cap j}$ and not in $\overline{\ppq_{i}}(N_{i \cap j})$. In order to define $\overline{\ppq_{i}}(\cdot)$, we first consider the set $\ppq_{i}$, which is recursively defined as follows:
    \begin{itemize}
        \item we initialize it as the empty set;
        \item at each following step, we pick some $j \in N_{i} \smin \{ i\}$ and we incorporate the functions within $N_{i \cap j}$ to $\ppq_{i}$.
    \end{itemize}
Given this set, we can define a map
\begin{equation*}
    \overline{\ppq_{i}}: \{N_{i \cap j}\}_{j \in N_{i} \smin \{i\}} \to 2^\me
\end{equation*}
such that, for each $j \in N_{i} \smin \{i\}$, $\overline{\ppq_{i}}(N_{i \cap j})$ stands for $\ppq_{i}$ at the step right before $j \in \mv$ is selected.
   
    \item For the internal energy $U$, we take the logarithm of the factors as energy terms and use
     \begin{equation*}
    U = - \sum_{(i,j) \in \me} \frac{1}{Z_{i \cap j}} \tr \left( \log f_{i,j} S_{i \cap j} \prod_{k \in N_{i \cap j}} \prod_{q \in  N_k} m_{k \cap q \to i \cap j} \right),
\end{equation*}
where $Z_{i \cap j}$ is a normalization constant
 \begin{equation*}
   Z_{i \cap j}= \tr \left(S_{i \cap j} \prod_{k \in N_{i \cap j}} \prod_{q \in  N_k} m_{k \cap q \to i \cap j} \right).
\end{equation*}
    \item For the Shannon entropy $S$, we use
   \begin{equation}
   \label{eq: approx entropy int}
\begin{split}
    S = &- \sum_{((i,j)) \in \mg} \frac{1}{\binom{|S_{i \cap j}|}{2}} \sum_{x_{i \cap j}} p_{i \cap j}(x_{i \cap j}) \log p_{i \cap j}(x_{i \cap j}) \\
    &- \sum_{(i,j) \in \me} W_{i,j} \sum_{x_i,x_j} p_{i,j}(x_i,x_j) \log p_{i,j}(x_i,x_j) \\
    &- \sum_{i \in \mv} C_i \sum_{x_i} p_i(x_i) \log p_i(x_i),
\end{split}
\end{equation}
where $W_{i,j}$ and $C_i$ are defined as in \eqref{eq: coefficients w and c},
the marginals $p_i$ are computed via \eqref{eq: approx marginals intersect}, and, omitting a normalization constant, 
\begin{equation}
    p_{i \cap j} (x_{i \cap j}) \propto S_{i \cap j} (x_{i \cap j}) \prod_{k \in N_{i \cap j}} \prod_{q \in  N_k \smin N_{i \cap j}}m_{k \cap q \to i \cap j} (x_k).
\end{equation}
Moreover, we can compute the two-variable marginals using the intersection neighborhoods, that is, omitting a normalization constant that ensures $p_{i,j}$ is a probability distribution, we have
\begin{equation*}
    p_{i,j}(x_i,x_j) \propto \tr_{(x_i,x_j)} \left( S_{i \cap j} \prod_{k \in N_{i \cap j}} \prod_{q \in  N_k} m_{k \cap q \to i \cap j} \right).
\end{equation*}

    \item For the partition function $Z$, we can use the equation for $U$ and $S$ above together with \eqref{eq: basic stat mech}. An alternative approach can also be established by following the discussion in Appendix \ref{sec: kcn inference}.
\end{itemize}

These approximate inference equations mimic their exact counterparts in Appendix \ref{sec: proof inference eq BOUNDED}. We simply remark that, in order to obtain the entropy formula \eqref{eq: approx entropy int}, we use that
\begin{equation*}
\begin{split}
    p(x) &= \prod_{\overline{i\cap j} \in \equi{\cap}} p_{i \cap j}(x_{i \cap j}) \prod_{i \in \mv} p_i(x_i)^{1-\left|\left\{\overline{k \cap q} \in \equi{\cap} | i \in \overline{k \cap q} \right\}\right|} \\
    &= \prod_{((i,j)) \in \mg} p_{i \cap j}(x_{i \cap j})^{\frac{1}{\binom{|S_{i \cap j}|}{2}}} \prod_{i \in \mv} p_i(x_i)^{1- \sum_{j \in N_i} \left( \frac{1}{|S_{i \cap j}|-1} \right)},
    \end{split}
\end{equation*}
where the first equality is \eqref{eq: probability kcn} and in the second equality we use the following relations:
\begin{equation}
\label{eq: counting equalities}
    \begin{split}
    &\sum_{j \in N_i \smin \{i\}} \left( \frac{1}{|S_{i \cap j}|-1} \right)-1 = \left|\left\{\overline{k \cap q} \in \equi{\cap} | i \in \overline{k \cap q} \right\}\right|-1,\\
&\sum_{k,q \in S_{i \cap j}} \frac{1}{\binom{|S_{k \cap q}|}{2}} = 1,
\end{split}
\end{equation}
which hold provided the loop bound is fulfilled.

\section{Comparison NIB-method and other message-passing schemes}
\label{sec: proof bounded claims}

\subsection{Proof of Claim \ref{claim: NIB vs kcn}}
    
    We start by showing sufficiency. To do so, we first note that it is enough to show a speed-up in inference. This is the case since the update complexity until convergence of the NIB-method is always bounded by that of the KCN-method. More specifically, this holds for the complexity of each update step since, by hypothesis, and for each $i \in \mv$ and $j \in N_i \smin \{i\}$ for which there exists some (eventually) non-uniform message, there is some $k \in N_i$ such that $N_{i \cap j} \subseteq N_{i \smin k}$. In fact, 
    the role of the (non-trivial) messages $m_{j \to i}^{(KCN)}$ in the KCN-method can be decomposed as follows:
     \begin{equation}
     \label{eq: nib decomposition of kcn}
       m_{j \to i}^{(KCN,t)} (x_j) \propto \prod_{\overline{j \cap k} \neq \overline{i \cap j}} m_{\overline{j \cap k} \to j}^{(t)}(x_j).
       \end{equation}
    Moreover, the number of update steps until convergence in the NIB-method is equal to that in the KCN-method.
    This is the case since the diameter of the NIB-associated hypernetwork is equal to that of each of the members of the family of directed tree networks that we can similarly associate to a network via the KCN-method. 
    
    We consider now the complexity of inference. To do so, take some $i_0 \in \mv$ such that, among all the sets $N_i$ with $i \in \mv$, the complexity of summing over $N_{i_0}$ is maximal. We consider two cases:
    If there is no $j \in \mv \cap \left( N_{i_0} \smin \{i_0\} \right)$ such that $N_{i_0} = N_{i_0 \cap j}$, then we do not need to sum over $N_{i_0}$ during inference in the NIB-method. Assume, hence, that there is some $j_0 \in \mv \cap \left( N_{i_0} \smin \{i_0\} \right)$ such that $N_{i_0} = N_{i_0 \cap j_0}$. In this scenario, and provided there exists some $k \in \mv \cap \left( N_{j_0} \smin N_{i_0 \cap j_0} \right)$, then $N_{i_0} \subsetneq N_{j_0}$ and we contradict the maximality of $N_{i_0}$. Conversely, if we have that $N_{j_0} \smin N_{i_0 \cap j_0} = \emptyset$ for any $j_0 \in \mv$ such that $N_{i_0} = N_{i_0 \cap j_0}$, then $N_{i_0 \cap j_0} = \mg$  by connectedness. This contradicts our hypothesis.
    Since we can argue along these lines for any $i_0 \in \mv$, the NIB-method never sums over the maxima (in terms of time complexity) of $\{N_i\}_{i \in \mv}$ during inference and, hence, it has a smaller time complexity than the KCN-method.
    
    To show necessity, we can argue by contrapositive. If there is some pair $i,j \in \mv$ such that $N_{i \cap j}= \mg$, then both the KCN and NIB methods have trivial update steps and the complexity of inference is that of summing over the whole network $\mg$ for both. Hence, the NIB-method achieves no speed-up.

  \subsection{Update step time complexity comparison between the KCN and NIB methods}
    
    \begin{claim}
    \label{claim: update speed-up nib vs kcn}
    If $r_0$ is a sharp loop bound, then the $r_0$-NIB-method has a smaller message update time complexity than the $r_0$-KCN-method if and only if, for each pair $i \in \mv$, $j \in \mv \cap \left( N_i \smin \{i\} \right)$ such that, among all the sets $N_{q \smin p}$ with $q \in \mv$ and $p \in \mv \cap \left( N_q \smin \{q\} \right)$, the complexity of summing over $N_{i \smin j}$ is maximal, there is no $k_{i \smin j} \in \mv$ such that $N_{i \smin j} = N_{i \cap k_{i \smin j}}$.
    \end{claim}
    
    To show sufficiency, we argue by contrapositive, assuming that, for some $N_{i \smin j}$ with maximal time complexity, there is some $k_{i \smin j} \in \mv$ such that $N_{i \smin j} = N_{i \cap k_{i \smin j}}$. This implies that we sum over $N_{i \smin j}$ in the update steps of the NIB-method and, hence, we have the same time complexity as in the KCN-method.
    
    To show necessity, we start by noting that, if we discard the trivial case (where $\mg=N_{i \cap j}$ for some $i,j \in \mv$ and there are no update steps in either method), then for each pair $i \in \mv$, $j \in \mv \cap \left( N_i \smin \{i\} \right)$, there exists some $q \in \mv$ and $p \in \mv \cap \left( N_q \smin \{q\} \right)$ such that $N_{i \cap j} \subseteq N_{q \smin p}$. Since, by hypothesis, we have $N_{i \cap j} \subsetneq N_{q \smin p}$ provided $N_{q \smin p}$ has maximal time complexity, we get that the time complexity of the update steps in the NIB-method is smaller than that of the KCN-method.

\subsection{Proof of Claim \ref{claim: NIB optimal}}

Before we prove the result, we ought to introduce a definition.

Given a network $\mg$, and a family of subsets of $\mg$, $M \equiv \{M_k\}_{k \in \mathcal K}$ with $M_k \subseteq \mg$ having the property that the endpoints of each edge that belongs to $M_k$ are also in $M_k$, we can define a BP-type algorithm as follows:
We associate a family of messages to each $M_k \in M$:
\begin{equation}
\label{eq: update ls-bp}
    \begin{split}
    &\{m_k^{(M,t)}\}_{k \in \mathcal K, t \geq 0}, \\
    &m_k: X \to \mathbb R_{\geq 0}.
    \end{split}
\end{equation}
These messages are uniformly initialized and updated, for each $t \geq 0$, according to
    \begin{equation}
    \label{eq: updates lsbp}
m_{k}^{(M, t+1)} (x_k) \propto  \tr_{\setminus x}  \biggl( \prod_{(i,j) \in M_k} f_{i,j}
\prod_{k_1 \in I(k)} m_{k_1}^{(M, t)} \biggr),
\end{equation}
where $x_k$ is some output variable in $\mg$, $I(k) \subseteq \mathcal K$, there is at least one message $m_{k_v}^{(M,t)}$ associated to each variable $v \in \mv \cap M_k$ that is not the output variable, and every message is used as input in the update of at least one other messages.
We call the family of such algorithms the \textbf{single variable BP algorithms} (\textbf{SVBP}). Note that standard BP, the KCN-method
and the NIB-method are SVBP algorithms. We call a SVBP algorithm \textbf{exact} provided it allows us to make exact inference.

Given the definition in the previous paragraph, we can now prove Claim \ref{claim: NIB optimal}.
We show the statement by reduction to the absurd.

Assume, in particular, that there is some SVBP algorithm $A_0$ whose message update time complexity is smaller than that of the NIB-method. Since the update complexity is smaller, there must exist some $i \in \mv$ and $j \in N_i \smin \{i\}$ 
such that $M_k \subsetneq N_{i \cap j}$ for all $k \in \mathcal K$. (This already discards the cases where NIB-method coincides with BP. We assume in what follows that the loop bound $r_0$ is strictly larger than zero.)

Since $A_0$ must communicate the information related to $i$ from $i$ to $j$, there must be some $M_i \in M$ such that $i \in M_i$ and, among the elements in $M$ that include $i$, $M_i$ is the set used in \eqref{eq: update ls-bp} which is closest (in the sense of the updates \eqref{eq: updates lsbp}) to some $M_j \in M$ such that $j \in M_j$. Because of this, and since the NIB-method is not BP, we have that $M_i \cap \left( N_{i \cap j} \smin \{i \}\right) \neq \emptyset$.
Nonetheless, since $M_i \subsetneq N_{i \cap j}$ by hypothesis, there must be some $e \in \me \cap  \left( N_{i \cap j} \smin M_i \right)$. In particular, there are at least two nodes $x_1,x_2 \in B(M_i) \cap N_{i \cap j}$, where $B(M_i)$ stands for the \textbf{boundary} of $M_i$, that is, the set of nodes in $M_i$ that are connected to edges outside of $M_i$.

By definition of the SVBP family, we can split the argument in two cases:
\begin{itemize}
    \item Both $x_1$ and $x_2$ act as input nodes in \eqref{eq: update ls-bp} when updating $m_{i}^{(M,t)}$. If this is the case, since we can pick $x_1$ and $x_2$ such that there is a path between them outside of $M_i$, there must be, for each of them, a sequence of neighborhoods $S_{x_1}, S_{x_2} \subseteq M$ used in \eqref{eq: update ls-bp} such that, in order for $A_0$ to be exact, the only intersection between members of different sequences must be the last one in each of them $L_{S_{x_i}} \in S_{x_i}$ for $i=1,2$, and the intersection between these two elements must be a single node $x_3$. Since $x_3$ is an input node for both neighborhoods, and to avoid destroying exactness by overcounting, the edges in $L_{S_{x_1}}$ connected to $x_3$ are not taken into account by $L_{S_{x_2}}$ (and vice versa).  This makes the algorithm inexact, and we have reached a contradiction. 
    \item One of them, say $x_1$, is an input node when updating $m_{i}^{(M,t)}$ and the other one is the output node. In this case, we can essentially argue like in the previous case. We should only take into account the fact that, in this scenario, it is possible for $x_3$ to be an input node of $L_{S_{x_1}}$ and the output node of $L_{S_{x_2}}$. If this is the case, we get a cycle in $A_0$ and we loose accuracy, reaching again a contradiction.
\end{itemize}

Since we have just argued that the message update time complexity cannot be smaller than that of the NIB-method, then the inference time complexity must be smaller. We can, however, reach a contradiction along similar lines.

\section{Inference in the interpolation between the KCN and NIB methods}
\label{sec: inference interpolation case}

We provide in this section the inference equations for the NIB-DIFF- and KCN-NIB-method.

\subsection{Inference in the NIB-DIFF-method}
\label{sec: inference Ni min Nj}

For brevity, in this section, we simply state three of the inference equations, and only consider the case where the loop bound is fulfilled. The missing cases can be derived following the ideas in other sections. Moreover, in contrast to the bounded NIB-method, we may have more messages in the NIB-DIFF-method. More specifically, we update non-trivially each message $m_{\overline{i \cap j} \to i}^{(t)}$, also those where $N_i \smin N_{i \cap j} = \emptyset$. This slight change follows from the fact that we intend to use the neighborhoods $N_{i \smin j}$ during inference and, without the change, some of the inference equations would not be well defined.

Given a network $\mg$ where the loop bound is fulfilled, the following inference equations are exact:

\begin{itemize}
   \item For the marginals $p_i(\cdot)$ with $i \in \mv$, we have that
   \begin{equation*}
    p_i(x_i) \propto \tr_{\smin x_i} \left( S_{i \smin j} \prod_{k \in N_{i \smin j}} \prod_{\overline{k \cap q} \not \subset N_{i \smin j}} m_{\overline{k \cap q} \to k} \right),
\end{equation*}
    where we omit a normalization constant that ensures $p_i$ is a probability distribution. We assume in this equation that, for each $i \in \mv$ and $j \in N_i$ we consider, there exists some $k \in N_i$ such that $\overline{i \cap j} \neq \overline{i \cap k}$. We can do so since, otherwise, the marginal inference is trivial
   $p_i \equiv m_{\overline{i \cap j} \to i}$.
    
    \item For the internal energy $U$, we take the logarithm of the functions in $\mg$ as energy terms and we get
    \begin{equation*}
\begin{split}
     U = &- \sum_{(i,j) \in \me} \frac{1}{Z_{i,j}} \tr \bigg( \log f_{i,j} S_{a_{i,j} \smin b_{i,j}} \\
     &\times \prod_{p_1 \in N_{a_{i,j} \smin b_{i,j}}} \prod_{\overline{p_1 \cap p_2} \not \subset N_{a_{i,j} \smin b_{i,j}}} m_{\overline{p_1 \cap p_2} \to p_1} \bigg)
     \end{split}
\end{equation*}
where we pick $a_{i,j}, b_{i,j} \in \mv$ such that $(i,j) \in  S_{a_{i,j} \smin b_{i,j}}$ \footnote{This only excludes the case where $\mg=N_{i \cap j}$ for $i,j \in \mv$, where the NIB-DIFF-method would essentially coincide with the NIB-method.}, and $Z_{i,j}$ is a normalization constant
\begin{equation*}
     Z_{i,j} \equiv \tr \left( S_{a_{i,j} \smin b_{i,j}} \prod_{p_1 \in N_{a_{i,j} \smin b_{i,j}}} \prod_{\overline{p_1 \cap p_2} \not \subset N_{a_{i,j} \smin b_{i,j}}} m_{\overline{p_1 \cap p_2} \to p_1} \right).
\end{equation*}

    \item For the partition function $Z$, we have that
     \begin{widetext}
\begin{equation*}
    Z = \frac{\prod_{i \in \mv} \prod_{j \in \equi{N_i \smin \{i\}}} \tr \left( S_{i \smin j} \prod_{k \in N_{i \smin j}} \prod_{\overline{k \cap q} \not \subset N_{i \smin j}} m_{\overline{k \cap q} \to k} \right)}{\prod_{\overline{i \cap j} \in \equi{\cap}} \tr \left( S_{i \cap j} \prod_{k \in N_{i \cap j}} \prod_{\overline{k \cap q} \neq \overline{i \cap j}} m_{\overline{k \cap q} \to k} \right)^{1- \sum_{k \in i \cap j} \left|\left\{ \overline{k \cap q} |  \overline{k \cap q} \neq \overline{i \cap j} \right\}\right|}}.
\end{equation*}
\end{widetext}
\end{itemize}

\subsection{Inference in the KCN-NIB-method}

As in the previous section, we simply state three of the inference equations, and only consider the case where the loop bound is fulfilled. In contrast to the update equations in the KCN-method, we include here more messages. More specifically, for each pair $i \in \mv$, $j \in N_i$ such that $N_{i \smin j}= \emptyset$, we include a message from $i$ to some \textbf{virtual} node $v_{i,j}$, for which we only need to assume $N_{i \smin v_{i,j}} = N_{i \cap j}$, and which we define and update in the KCN fashion as if $v_{i,j}$ was an actual node. These messages are only relevant for the computation of marginals and are ignored in the other inference equations.

Given a network $\mg$ where the loop bound is fulfilled, the following inference equations are exact:

\begin{itemize}
    \item For the marginals $p_i(\cdot)$ with $i \in \mv$, omitting a normalization constant that ensures $p_i$ is a probability distribution, we have that
    \begin{widetext}
     \begin{equation*}
        p_i(x_i) \propto
\prod_{j \in N_i \smin \{i\}} \left(m_{i \to j}^{\text{(KCN)}}(x_i)\right)^{\left( \left(|S_{i \cap j}|-1\right) \left( \sum_{j \in N_i} \left( \frac{1}{|S_{i \cap j}|-1} \right)-1 \right) \right)^{-1}}.
    \end{equation*}
    \end{widetext}
    \item For the internal energy $U$, we take the logarithm of the functions in $\mg$ as energy terms and we get
  \begin{equation*}
    U = - \sum_{(i,j) \in \me} \frac{1}{Z_{i \cap j}} \tr \left( \log f_{i,j} S_{i \cap j} m_{i \to j}^{\text{(KCN)}} \prod_{k \in N_{i \cap j} \smin \{ i\}} m_{k \to i}^{\text{(KCN)}} \right),
\end{equation*}
where $Z_{i \cap j}$ is a normalization constant
 \begin{equation*}
   Z_{i \cap j} \equiv \tr \left(S_{i \cap j} m_{i \to j}^{\text{(KCN)}} \prod_{k \in N_{i \cap j} \smin \{ i\}} m_{k \to i}^{\text{(KCN)}} \right).
\end{equation*}
    
    \item For the partition function $Z$, we have that
     \begin{widetext}
 \begin{equation*}
 Z = \frac{\prod_{((i,j)) \in \mg} \tr \left(S_{i \cap j} m_{j \to i}^{\text{(KCN)}} \prod_{k \in N_{i \cap j} \smin \{j\}} m_{k \to j}^{\text{(KCN)}} \right)^\frac{1}{\binom{|S_{i \cap j}|}{2}}}{\prod_{i \in \mv_{\text{NT}}} \tr \left( \prod_{j \in N_i \smin \{i\}} \left(m_{i \to j}^{\text{(KCN)}}\right)^{\left( \left(|S_{i \cap j}|-1\right) \left( \sum_{j \in N_i \smin \{i\}} \left( \frac{1}{|S_{i \cap j}|-1} \right)-1 \right) \right)^{-1}} \right)^{ \sum_{j \in N_i \smin \{i\}} \left( \frac{1}{|S_{i \cap j}|-1} \right)-1}},
  \end{equation*}
 \end{widetext}
 where 
 \begin{equation*}
     \mv_{\text{NT}} \equiv \{ i \in \mv : |\equi{N_i \smin \{i\}}| \geq 2 \}.
 \end{equation*}
\end{itemize}

For brevity, and since the other equations are at most as difficult as this one, we only show the inference equation for the partition function $Z$, which can be derived via the following chain of equalities:
\begin{widetext}
  \begin{equation}
  \label{eq: derivation Z kcn}
     \begin{split}
    &\frac{\prod_{((i,j)) \in \mg} \tr \left(S_{i \cap j} m_{j \to i}^{\text{(KCN)}} \prod_{k \in N_{i \cap j} \smin \{j\}} m_{k \to j}^{\text{(KCN)}} \right)^\frac{1}{\binom{|S_{i \cap j}|}{2}}}{\prod_{i \in \mv_{\text{NT}}} \tr \left( \prod_{j \in N_i \smin \{i\}} \left( m_{i \to j}^{\text{(KCN)}}\right)^{\left( \left(|S_{i \cap j}|-1\right) \left( \sum_{j \in N_i \smin \{i\}} \left( \frac{1}{|S_{i \cap j}|-1} \right)-1 \right) \right)^{-1}} \right)^{ \sum_{j \in N_i \smin \{i\}} \left( \frac{1}{|S_{i \cap j}|-1} \right)-1}} \\
     &=\frac{\prod_{\overline{i \cap j}} \tr \left(S_{i \cap j} m_{j \to i}^{\text{(KCN)}} \prod_{k \in N_{i \cap j} \smin \{j\}} m_{k \to j}^{\text{(KCN)}} \right)}{\prod_{i \in \mv_{\text{NT}}} \tr \left( \prod_{j \in (N_i \smin \{i\})/ \sim} \left( m_{i \to j}^{\text{(KCN)}}\right)^{\left( \sum_{j \in N_i \smin \{i\}} \left( \frac{1}{|S_{i \cap j}|-1} \right)-1\right)^{-1}} \right)^{ \sum_{j \in N_i \smin \{i\}} \left( \frac{1}{|S_{i \cap j}|-1} \right)-1}} \\
     &=\frac{\prod_{\overline{i \cap j}} \tr \left(S_{i \cap j} m_{j \to i}^{\text{(KCN)}} \prod_{k \in N_{i \cap j} \smin \{j\}} m_{k \to j}^{\text{(KCN)}} \right)}{\prod_{i \in \mv_{\text{NT}}} \tr \left( \prod_{j \in (N_i \smin \{i\})/ \sim} \left( m_{i \to j}^{\text{(KCN)}}\right)^{\left(|\{j \in N_i\smin \{i\}\}/\sim|-1\right)^{-1}} \right)^{|\{j \in N_i \smin \{i\}\}/\sim|-1}} \\
     &=\frac{\prod_{\overline{i \cap j}} \tr \left(S_{i \cap j} n_{j \to i}^{\text{(KCN)}} \prod_{k \in N_{i \cap j} \smin \{j\}} n_{k \to j}^{\text{(KCN)}} \right)}{\prod_{i \in \mv_{\text{NT}}} \tr \left( \prod_{j \in (N_i \smin \{i\})/ \sim} \left( n_{i \to j}^{\text{(KCN)}}\right)^{\left(|\{j \in N_i\smin \{i\}\}/\sim|-1\right)^{-1}} \right)^{|\{j \in N_i \smin \{i\}\}/\sim|-1}} \\
     &= \frac{\prod_{\overline{i \cap j}} Z}{\prod_{i \in \mv_{\text{NT}}} \tr \left( \prod_{j \in (N_i \smin \{i\})/ \sim} \left( \prod_{k \in (N_i \smin N_j)/ \sim} n_{\overline{i \cap k} \to i} \right)^{\left(|\{j \in N_i\smin \{i\}\}/\sim|-1\right)^{-1}} \right)^{|\{j \in N_i\smin \{i\}\}/\sim|-1}} \\
     &= \frac{\prod_{\overline{i \cap j}} Z}{\prod_{i \in \mv_{\text{NT}}} \tr \left( \prod_{k \in (N_i \smin \{i\})/ \sim} n_{\overline{i \cap k} \to i} \right)^{|\{j \in N_i\smin \{i\}\}/\sim|-1}} \\
     &= \frac{\prod_{\overline{i \cap j}} Z}{\prod_{i \in \mv_{\text{NT}}} Z^{|\{j \in N_i\smin \{i\}\}/\sim|-1}} \\
     &= Z,
     \end{split}
 \end{equation}
 \end{widetext}

 where, in the first equality, we use the second equality in \eqref{eq: counting equalities} together with the fact that, for all $i,j,k \in \mg$, $m_{i \to j}^{\text{(KCN)}} = m_{i \to k}^{\text{(KCN)}}$ provided $S_{i \cap j}=S_{i \cap k}$. 
In the second equality, we use the first equality in \eqref{eq: counting equalities}.
In the third equality we use the notation $n_{i \to j}^{\text{(KCN)}}$ to refer to $m_{i \to j}^{\text{(KCN)}}$ without the normalization constants and note that, given the powers in the denominator, each normalization constant $M_{i \to j}^{\text{(KCN)}}$ in the upper part gets cancelled with one in the lower part. In the fourth equality, we decompose the messages $n_{i \to j}^{\text{(KCN)}}$ into their constituent intersection messages (that is, into the product of the unnormalized messages in the NIB-method that carry the same information as $n_{i \to j}^{\text{(KCN)}}$). In the last equality, we use a result analogous to \cite[Theorem 9.1]{wilson1979introduction} for hypergraphs.
\end{appendix}
\end{document}